\documentclass{aastex701}

\begin{document}

\title{Local Interstellar Flow Parameters from the First Intersection of IMAP-Lo's Parameter Tubes }

\author[orcid=0000-0002-3737-9283,gname=Nathan A., sname='Schwadron']
{Nathan A. Schwadron} 
\affiliation{Los Alamos National Laboratory, PO Box 1663, Los Alamos, NM 87545, USA}
\affiliation{University of New Hampshire, Durham, NH 03824, USA}
\affiliation{Department of Astrophysical Sciences, Princeton University, Princeton, NJ 08544, USA}
\email{nschwadron@lanl.gov}

\author[orcid=0000-0002-3093-458X,gname=Mitchell M.,sname='Shen']
{Mitchell M. Shen}
\affiliation{Department of Astrophysical Sciences, Princeton University, Princeton, NJ 08544, USA}
\email{mitchellshen@princeton.edu}  

\author[orcid=0000-0001-9316-0553,gname=Fatemeh,sname='Rahmanifard']
{Fatemeh Rahmanifard}
\affiliation{University of New Hampshire, Durham, NH 03824, USA}
\email{f.rahmanifard@unh.edu} 

\author[orcid=0000-0002-2745-6978,gname=Eberhard,sname='MAbius']
{Eberhard M\"obius}
\affiliation{University of New Hampshire, Durham, NH 03824, USA}
\email{eberhard.moebius@unh.edu} 

\author[orcid=0000-0002-1160-7022,gname=Hafijul,sname='Islam']
{Hafijul Islam}
\affiliation{University of New Hampshire, Durham, NH 03824, USA}
\email{hafijul.islam@unh.edu} 

\author[orcid=0000-0002-4103-9935,gname=Jonathan S.,sname='Bower']
{Jonathan S. Bower}
\affiliation{University of New Hampshire, Durham, NH 03824, USA}
\email{jonathan.bower@unh.edu} 

\author[orcid=0000-0003-3957-2359,gname=Maciej,sname='Bzowski']
{Maciej Bzowski}
\affiliation{Space Research Centre PAS (CBK PAN), Bartycka 18A, 00-716 Warsaw, Poland}
\email{bzowski@cbk.waw.pl}  

\author[orcid=0000-0003-2134-3937,gname=Eric R.,sname='Christian']
{Eric R. Christian}
\affiliation{Heliophysics Science Division, NASA Goddard Space Flight Center, Greenbelt, MD 20771, USA}
\email{eric.r.christian@nasa.gov} 

\author[gname=Ken,sname='Fairchild']{Ken Fairchild}
\affiliation{University of New Hampshire}
\email{Ken.Fairchild@unh.edu}

\author[orcid=0000-0002-6817-1039,gname=Herbert O, sname='Funsten']
{Herbert O. Funsten} 
\affiliation{Los Alamos National Laboratory, PO Box 1663, Los Alamos, NM 87545, USA}
\email{hfunsten@lanl.gov}

\author[orcid=0000-0003-4101-7901,gname=Stephen A.,sname='Fuselier']
{Stephen A. Fuselier}
\affiliation{Southwest Research Institute, San Antonio, TX 78238, USA}
\affiliation{University of Texas at San Antonio, San Antonio, TX 78249, USA}
\email{stephen.fuselier@swri.org}

\author[orcid=0000-0003-2425-3793,gname=Andr\'e,sname='Galli']
{Andr\'e Galli}
\affiliation{Space Research and Planetary Sciences, Physics Institute, University of Bern, Bern, Switzerland}
\email{andre.galli@unibe.ch}  

\author[orcid=0000-0002-9237-7088,gname=Jonathan,sname='Gasser']
{Jonathan Gasser}
\affiliation{Southwest Research Institute, San Antonio, TX 78238, USA}
\email{jonathan.gasser@swri.org} 

\author[orcid=0000-0001-9979-2164,gname=Matina,sname='Gkioulidou']
{Matina Gkioulidou}
\affiliation{Applied Physics Laboratory, The Johns Hopkins University, Laurel, MD 20723, USA}
\email{matina.gkioulidou@jhuapl.edu} 

\author[gname=David,sname='Heirtzler']{David Heirtzler}
\affiliation{University of New Hampshire}
\email{David.Heirtzler@unh.edu}

\author[orcid=0000-0002-6569-3800,gname=Izabela ,sname='Kowalska-LeszczyAska']
{Izabela Kowalska-Leszczyska}
\affiliation{Space Research Centre PAS (CBK PAN), 5 Bartycka 18A, 00-716 Warsaw, Poland}
\email{ikowalska@cbk.waw.pl}

\author[orcid=0000-0002-5204-9645,gname=Marzena,sname='Kubiak']
{Marzena A. Kubiak}
\affiliation{Space Research Centre PAS (CBK PAN), 5 Bartycka 18A, 00-716 Warsaw, Poland}
\email{mkubiak@cbk.waw.pl}

\author[orcid=0000-0001-6160-1158,gname=David J.,sname='McComas']
{David J. McComas}
\affiliation{Department of Astrophysical Sciences, Princeton University, Princeton, NJ 08544, USA}
\email{dmccomas@princeton.edu}  

\author[orcid=0000-0001-6286-5809,gname=Jonathan T.,sname='Niehof']
{Jonathan T. Niehof}
\affiliation{University of New Hampshire, Durham, NH 03824, USA}
\email{Jonathan.Niehof@unh.edu} 

\author[orcid=0000-0002-8111-1444,gname=Jamie S.,sname='Rankin']
{Jamie S. Rankin}
\affiliation{Department of Astrophysical Sciences, Princeton University, Princeton, NJ 08544, USA}
\email{jsrankin@princeton.edu}  

\author[orcid=0000-0002-4173-3601,gname=Justyna,sname='SokA3A']
{Justyna M. Sok\'o\l}
\affiliation{Southwest Research Institute, San Antonio, TX 78238, USA}
\email{justyna.sokol@swri.org}

\author[orcid=0000-0002-9033-0809,gname=PaweA,sname='Swaczyna']
{Pawe\l{} Swaczyna}
\affiliation{Space Research Centre PAS (CBK PAN), Bartycka 18A, 00-716 Warsaw, Poland}
\email{pswaczyna@cbk.waw.pl} 

\author[orcid=0000-0003-2685-9801,gname=Jamey R.,sname='Szalay']
{Jamey R. Szalay}
\affiliation{Department of Astrophysical Sciences, Princeton University, Princeton, NJ 08544, USA}
\email{jszalay@princeton.edu}  

\author[gname=Carol,sname='Weaver']
{Carol Weaver}
\affiliation{University of New Hampshire, Durham, NH 03824, USA}
\email{Carol.Weaver@unh.edu} 

\author[orcid=0000-0002-2603-1169,gname=Peter,sname='Wurz']
{Peter Wurz}
\affiliation{Space Research and Planetary Sciences, Physics Institute, University of Bern, Bern, Switzerland}
\email{peter.wurz@unibe.ch}

\author[orcid=0000-0002-2603-1169,gname=Gary P.,sname='Zank']
{Gary P. Zank}
\affiliation{Department of Space Science, University of Alabama in Huntsville, Huntsville, AL 35899, USA}
\email{gary.zank@uah.edu}

\author[orcid=0000-0001-7240-0618,gname=Eric J.,sname='Zirnstein']
{Eric J. Zirnstein}
\affiliation{Department of Space Science, University of Alabama in Huntsville, Huntsville, AL 35899, USA}
\email{eric.zirnstein@uah.edu}






\correspondingauthor{Nathan Schwadron}
\email{nschwadron@lanl.gov}


\begin{abstract}

The Sun's motion through the interstellar medium creates a flow of interstellar neutral (ISN) atoms through the heliosphere. ISN He, due to its high universal abundance and relatively low ionization rate, is the most abundant of the interstellar species near 1 au and ideal for flow parameter determination.   The Interstellar Boundary Explorer (IBEX) measurements of ISN He flow parameters (speed, temperature, and direction) yielded a tube in 4D parameter space -- narrow in cross-section but highly extended along  one parameter axis (e.g., ecliptic longitude direction). This ``4D parameter tube'' results in large systematic uncertainties, a direct consequence of IBEX-Lo's fixed viewing orientation on the spacecraft. On the Interstellar Mapping and Acceleration Probe (IMAP), the articulation of the IMAP-Lo boresight using its pivot platform enables multiple viewing orientations of the ISN flow for significant systematic uncertainty reduction. We provide first results that definitively intersect ISN parameter tubes for elongation angles  $79^\circ$, 94$^\circ$, and 109$^\circ$, resulting in precise interstellar parameters: speed $26.37 \pm 0.82$ km s$^{-1}$, ecliptic longitude direction $74.85^\circ \pm 0.96^\circ$, ecliptic latitude direction $-5.212^\circ \pm 0.035^\circ$, and temperature $7740^{+770}_{-730}$ K. The  inferred flow of the Very Local Interstellar Medium is not consistent with either the Local Interstellar Cloud or the G-Cloud, but rather an intermediate state. IMAP is now positioned to study the detailed physics of this complex, nearby interstellar region. By resolving and understanding its physics, we determine how the heliosphere responds to the local interstellar flow, and how it may evolve in time.


\end{abstract}



\section{Introduction}

The Interstellar Mapping and Acceleration Probe (IMAP) mission investigates two fundamental and interrelated questions  at the heart of Heliophysics today: (1) How are energetic particles accelerated in interplanetary space, and (2) How does the solar wind interact  with the interstellar medium \cite[LISM,][]{McComas:2025}? The IMAP-Lo instrument  \cite[]{Schwadron:2025} provides accurate measurements of interstellar flow parameters by measuring and mapping interstellar neutral atoms over a range of elongation angles (the elongation is the boresight viewing angle relative to the Sun). The interstellar flow parameters (flow direction, temperature and speed) set the outer boundary conditions of the interstellar interactions. The inner boundary conditions are imposed by the solar wind, and together the inner and outer boundary conditions set up the structure and physical properties of our heliosphere. In particular, the velocity vector of the LISM relative to the Sun  is a central  quantity that regulates the interactions between the Very Local Interstellar Medium (VLISM) and the heliosphere \cite[]{Szalay:2026}.  

Interstellar neutral (ISN) gas penetrates deeply into the heliosphere due to the relative motion of the Sun with respect to its surroundings in the VLISM. Unlike the ions within the VLISM, which are guided by the plasma including the magnetic field, and deflected around the heliopause, the interstellar neutral atoms are affected by the plasma through charge-exchange and occasional collisions. The collision distance scales in the interstellar medium are 100's of astronomical units (au), which indicates that most neutral atoms can travel deeply into the heliosphere unimpeded by the solar wind. Interstellar He atoms are particularly important for the determination of interstellar parameters since He has a relatively high universal abundance, second only to H, and He has a high first ionization potential, which means that ionization rates of He are quite low (typically $\sim$ $1\times 10^{-7}$ s$^{-1}$ at 1 au, and fall off roughly with the inverse square of distance from the Sun). With such a low ionization rate, the majority of He atoms from the interstellar medium travel well inside of 1 au (typically to $\sim$0.5 au), and their measurement by IMAP presents  the best proxy for the determination of VLISM flow parameters. 

The Local Interstellar Medium (LISM) consists of a low-density plasma within a cavity associated with the Local Bubble,
which extends more than 100 pc from the Sun \cite[]{Frisch:1983}, i.e., much further away than the VLISM. There are numerous interstellar clouds within this region \cite[]{redfield:2008}. In the VLISM nearby the heliosphere, the local interstellar flow has a velocity vector directed away from the center of the Loop I superbubble \cite[]{Frisch:2013}. The nearest shell (S1) to the heliosphere is centered $\sim 78$ pc away with a radius of $\sim 82$ pc \cite[]{Wolleben:2007}. Nearest to the heliosphere are a group of clouds with temperatures ranging from 300 -- 13,000 K and proton densities in the range $\sim$ 0.05 to 0.1 cm$^{-3}$. 

\begin{figure*}[ht!]
\centering
\includegraphics[width=0.6\columnwidth]{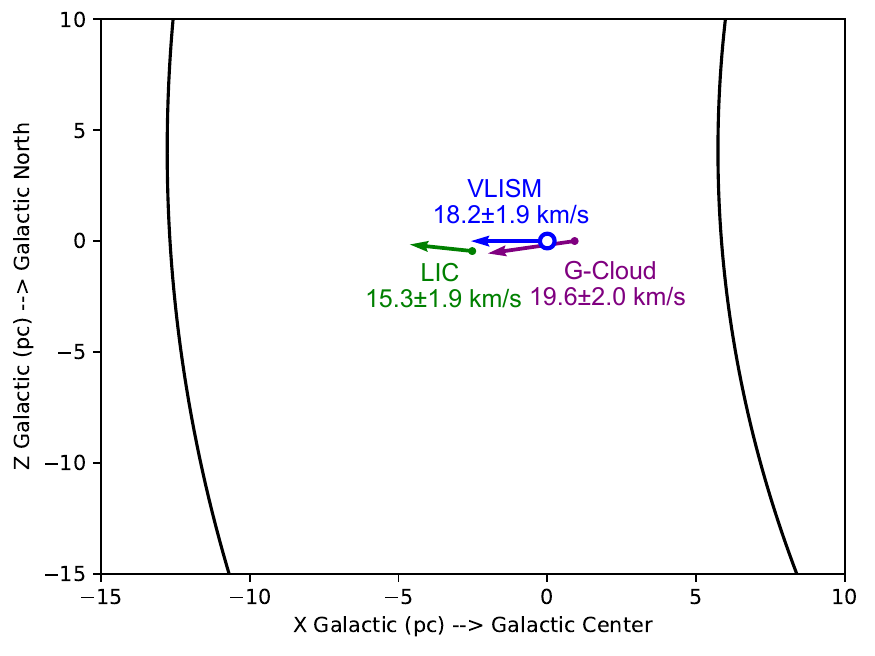}
\caption{
Loop I S1 inner and outer shells \cite[]{Wolleben:2007} together with the velocity vectors of the LIC cloud  \cite[]{redfield:2008} (green), the G-cloud (purple) and the VLISM (blue) \cite[]{schwadron:2022}. The vectors are referenced to the Local Standard of Rest frame \cite[]{Schonrich:2010} in the galactic medium. The locations of the clouds are indicated based on the central coordinates and the distance to the closest star in the group of star sight lines used for cloud velocity determination \cite{redfield:2008}. 
 }.
\label{fig:clouds}
\end{figure*}

Figure \ref{fig:clouds} shows the Loop I S1 inner and outer shells \cite[]{Wolleben:2007} together with the velocity vectors of the Local Interstellar Cloud (LIC) cloud (green), the G-cloud (purple) and the VLISM (blue). The cloud velocities are from \cite[]{redfield:2008}, the VLISM velocity is from \cite{schwadron:2022}, and each of these velocity vectors is referenced to the Local Standard of Rest (LSR) frame \cite[]{Schonrich:2010} in the galactic medium. The uncertainties in each of the cloud and VLISM velocities take into account both uncertainties in the LSR frame and the velocities themselves. The current value of the speed of the neutrals in the  VLISM \cite[$25.99^{+1.86}_{-1.76}$ km s$^{-1}$ in the Sun's rest frame,][]{schwadron:2022} has a large uncertainty due to the degeneracy in the ISN parameters. The locations of the clouds are indicated based on the central coordinates and the distance to the closest star in the group of star sight lines used for cloud velocity determination \cite[]{redfield:2008} (there were 79 sight lines used for the LIC cloud, and 21 sight lines for the G-Cloud). 

It is remarkable that the VLISM could be in an intermediate state between the LIC and the G-Cloud. As noted by \cite{redfield:2008}, ``the heliosphere appears to be
located in a transition zone between the LIC and G Cloud.'' This notion of a transition region, or a mixing region, was shown as very likely by \cite{swaczyna:2022clouds} based on the ISN parameters derived from IBEX observations and the observed asymmetry in the interstellar He distribution. While significant uncertainties remain, the data reveals that the speeds decrease with distance in the Loop I shell, which is consistent with a decelerating flow in this medium \cite[]{Frisch:2013}. Cementing this current understanding of the VLISM requires that we break the observational degeneracy resulting from the fixed viewing of the interstellar flow on IBEX, which is the subject of this paper. 

A  consolidation of interstellar parameters by \cite{schwadron:2022} revealed  convergence to a defined set of parameters from a wide array of observations including: neutral gas measurements by Ulysses \cite[]{witte:2004a, Wood:2015} and IBEX \cite[]{McComas:2015, bzowski:2015, schwadron:2015-isnhe, swaczyna:2018}, the UV backscatter measurements by EUVE \cite[]{Vallerga:2004} and Prognoz 6 \cite[]{lallement:2004a}, and the measurements of pickup ions by ACE \cite[]{Gloeckler:2004} STEREO \cite[]{taut:2018a, bower:2019}. Further interstellar temperature and speed measurements are determined from UV absorption based on Haute-Provence \cite[]{lallement:1992a} and Hubble \cite[]{Linsky:1993}.  

The IBEX-Lo ISN measurements, because they are taken at roughly the same location around the Sun each year, yield a direct relationship between the interstellar flow longitude direction, latitude direction, the speed, and temperature via the neutral atom trajectory equation\cite[]{lee:2012}. This relation yields a large uncertainty in the longitude direction, which carries over into all other interstellar parameters \cite[e.g.,][]{schwadron:2022, swaczyna:2022b}.  An independent determination of the ISN longitude is needed to remove systematic uncertainties associated with all interstellar parameters.  The pivot platform on IMAP-Lo was designed to track the interstellar flow over an extended longitude range  of spacecraft positions around the Sun  and thereby vastly increases the interstellar neutral atom observation period throughout the year by providing an array of different viewing orientations to observe the interstellar flow. 

In this paper, we  briefly introduce IMAP-Lo and the observations in \S 2. In \S 3, we discuss the methodology used to reduce systematic uncertainties by crossing the 4D parameter tubes. In \S 4, we discuss the key observations required to define interstellar parameters: longitude and spin phase distributions at the pivot platform angles studied. We infer VLISM flow parameters in \S 5, and discuss the implications of these observations in \S 6.  Conclusions are provides in \S 7 and there are three appendices in the paper. Appendix A discusses the correction factors applied to the peak spin phase rates as a function of spacecraft longitude. Appendix B describes the angular shifts associated with electronic timing lags in the instrument, the observatory, and the angular shift due to the alignment between the star sensor and the instrument boresight. Appendix A details the corrections associated with secondary neutrals, conversion to phase-space-density (PSD), ionization loss, and the sputtering efficiency. Appendix B details the spin phase peak corrections applied to the Histogram data, and Appendix C describes the mounting offsets that are accounted for in the search for VLISM parameters. 
 Appendix D describes Ecliptic J2000 frame used for all attitude, physical position and velocity data. Consistency checks were performed to ensure rigorous use of the Ecliptic J2000 frame. 

\section{The IMAP-Lo Instrument}

\begin{figure}
\centering
\includegraphics[width=0.8\columnwidth]{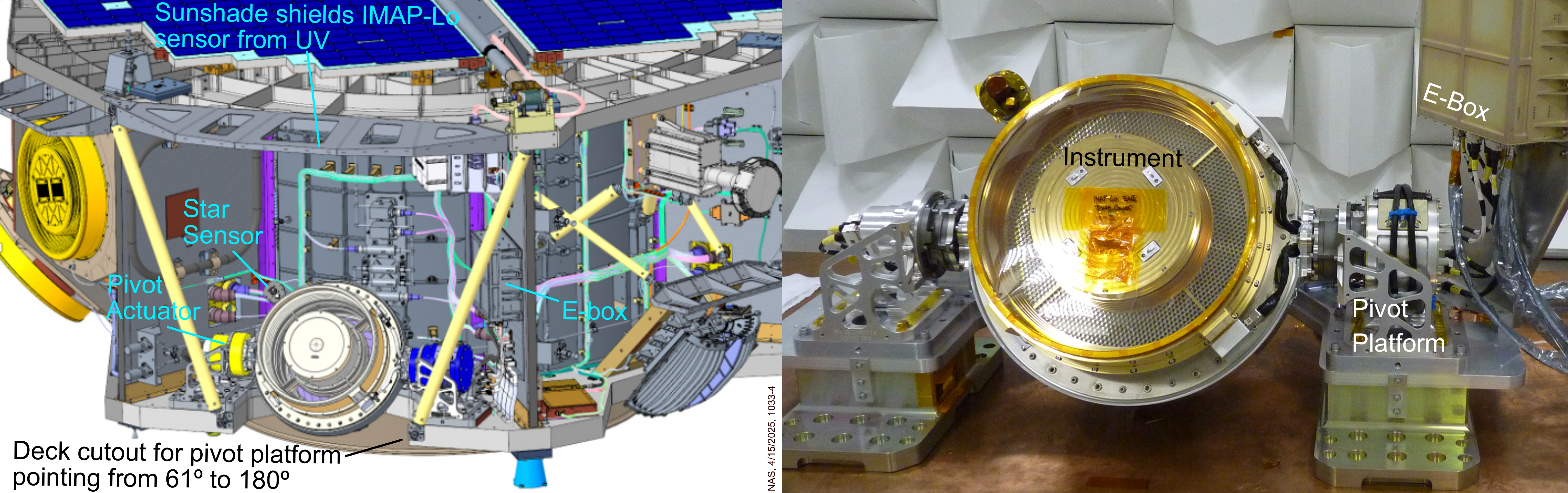} 
\caption{
(Left) IMAP-Lo mounted on the IMAP spacecraft, illustrating the locations of the electronics box (E-Box), the actuator on the pivot platform mechanism (PPM), the sunshade, and the star sensor; adapted from \cite{Schwadron:2025}. (Right) IMAP-Lo hardware during testing, showing the PPM, sensor assembly, and E-Box.
 }
\label{fig:mainComponents}
\end{figure}

IMAP-Lo is a single-pixel camera designed to image energetic neutral atoms (ENAs) at energies below 1 keV and, importantly, to measure interstellar neutral atoms (ISNs) from the incoming ISN flow. The instrument builds largely on the heritage design of IBEX-Lo \cite[]{Fuselier:2009b} and includes a key new capability: the Pivot Platform Mechanism (PPM). Unlike IBEX-Lo, the PPM allows IMAP-Lo to track the ISN flow over much of the year, significantly improving its ability to observe and characterize interstellar neutral atoms while also enabling ENA measurements from a range of pivot angles. The IMAP orbit near L1 enables IMAP-Lo observations that are free from magnetospheric background, which greatly reduced IBEX-Lo observation times. The primary components of the IMAP-Lo sensor, the PPM, and the E-Box are shown in Figure \ref{fig:mainComponents}. 

In this section, we discuss the measurement concept and sputtering by He in \S 2.1. We provide an overview of operations in the first year in \S 2.2, and the data products used for analysis of ISN He in \S 2.3. 

\subsection{Measurement Concept and Sputtering by He}

\begin{figure}
\centering
\includegraphics[width=0.6\columnwidth]{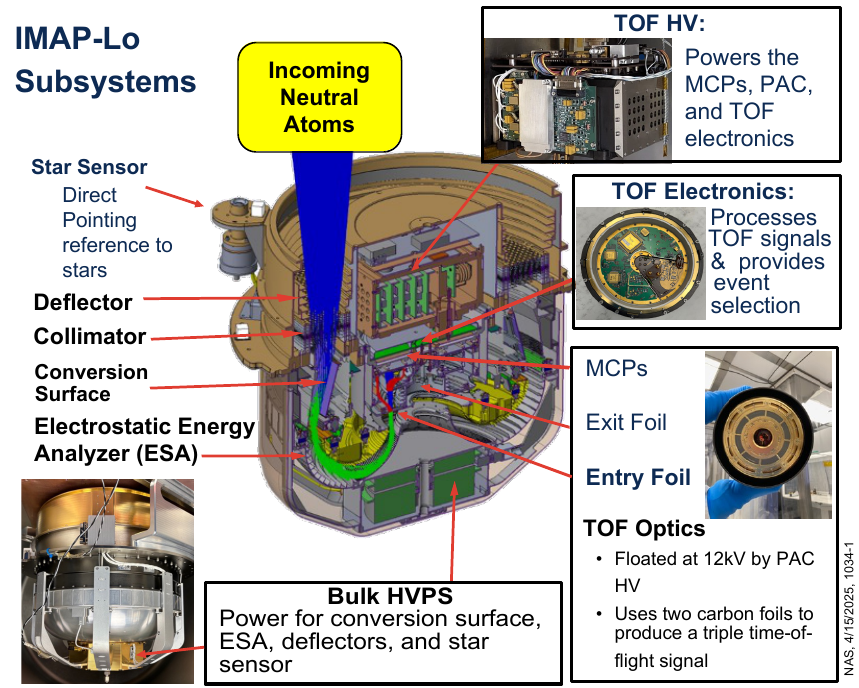} 
\caption{
Cutaway view adapted from \cite[]{Schwadron:2025} of IMAP-Lo with SIMION ray-tracing overlaid on the left side. Incident energetic neutral atoms (ENAs) are shown in blue, negative ions produced at the conversion surface (CS) are shown in green, particles transmitted through the carbon foils in the TOF section are shown in blue, and secondary electrons emitted from the foils are shown in red. Major subsystems are labeled, including the star sensor, deflector, collimator, CS, electrostatic analyzer (ESA), bulk high-voltage power supply (Bulk HVPS), TOF optics, entry and exit foils, microchannel plates, TOF electronics, and TOF high-voltage power supply (TOF HVPS).
 }
\label{fig:subsystems}
\end{figure}

The IMAP-Lo measurement concept and primary subsystems are shown in Figure \ref{fig:subsystems}. Neutral atoms, including interstellar neutral atoms from the inflowing ISN population and ENAs, enter the instrument through the large annular aperture and are collimated along the blue trajectories. A deflector system prevents ambient ions and electrons from entering the sensor, ensuring that the measured particles originate  from neutral populations. After geometric collimation, the atoms strike a diamond-like carbon (DLC) conversion surface  at a shallow angle of approximately $15^\circ $ \cite[]{scheer:2006,sokol:2024}.

At the conversion surface, a fraction of the incident neutrals are transformed into negative ions and are transmitted into the electrostatic analyzer (ESA), shown by the green trajectories. This occurs either by direct charge conversion, in which the neutral atom receives an electron from the surface, or by sputtering of atoms and molecules from the surface. In the case of ISN He, because it is a  noble element, the products are predominantly ions  sputtered from the mono-layers of molecules (predominantly hydroxyl group molecules) present on top of the conversion surface. The singly charged negative ions sputtered off the conversion surfaces are then accelerated into the toroidal electrostatic analyzer (ESA), which selects particles by energy-per-charge. After exiting the ESA, the ions are accelerated and electrostatically focused into the time-of-flight (TOF) subsystem, floated to approximately $12$ kV. Through this sequence, IMAP-Lo converts incoming neutral atoms into charged particles suitable for energy and mass analysis, enabling measurements of both ENAs and the ISN flow. 

\begin{figure}
\centering
\includegraphics[width=0.8\columnwidth]{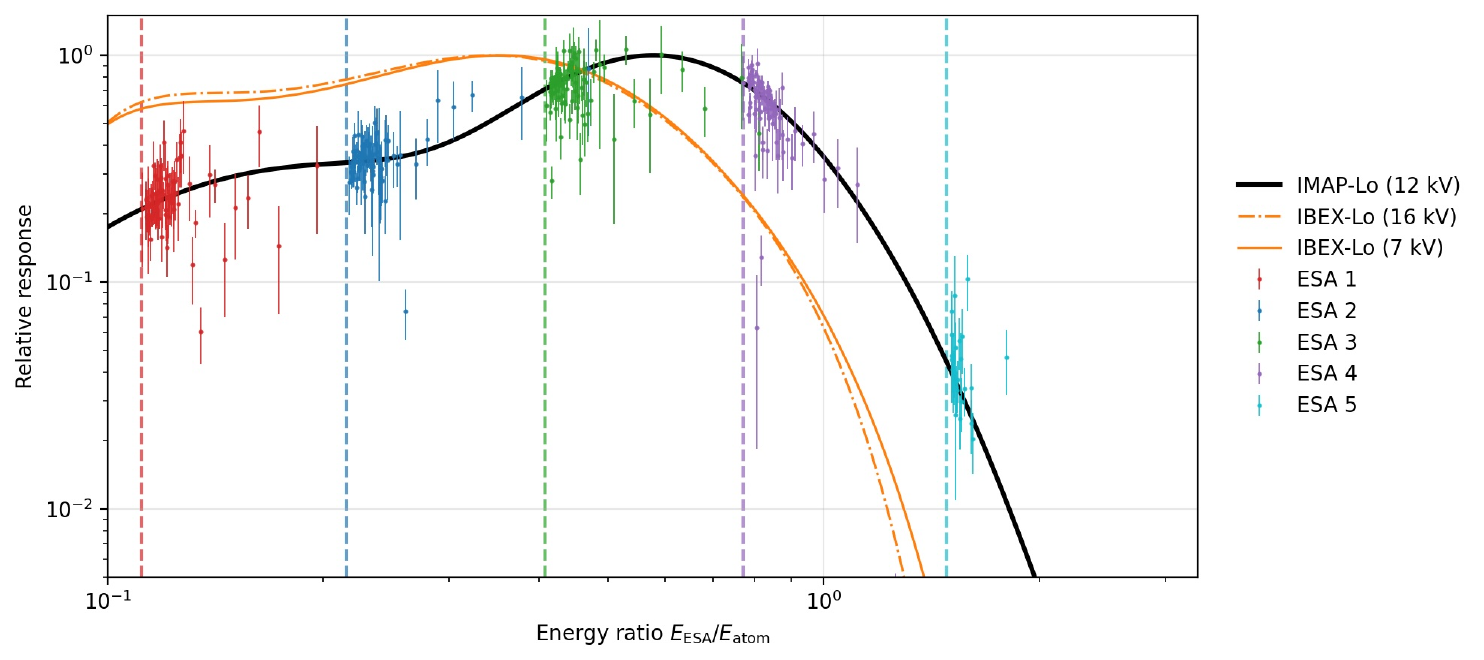} 
\caption{
The response function of IMAP-Lo due to sputtering by atoms such as He. ISN He atoms strike the conversion surface and produce H$^-$ and O$^-$. The relative efficiency of H$^-$ sputtering has been obtained as a function of the ESA central energy relative to the incident ISN energy of He atoms. The normalized fit response curve for IMAP-Lo (black) is shown together with that from  IBEX-Lo (dashed orange curve when the IBEX-Lo PAC was at 16 kV early in the mission, and solid orange when the PAC was at 7 kV after July 2012). 
The response curve for IBEX-Lo shown here was calibrated in flight while the IBEX-Lo TOF detector \cite[]{schwadron:2022, swaczyna:2023}.
The data points represent the ratio of the count rates to the average incoming flux. 
 }
\label{fig:sputtering}
\end{figure}

We have performed an in-flight calibration to determine the sputtering efficiency response from the conversion surface (Figure \ref{fig:sputtering}). Several takeaways result from the analysis, as described here:
\begin{itemize} 
\item The sputtering response function of IMAP-Lo is significantly higher in energy than that of IBEX-Lo. This does not affect the central energy steps, which are controlled by the ESA. However, the  response to species such as He, which predominantly cause sputtering, moves up in energy. 
\item The precise reason for this difference is still being investigated, although the difference must somehow reflect the conversion surface behavior, possibly due to ultra cleanliness of the instrument  and lack of hydrocarbon contamination on the conversion surface. 
\item The sputtering products observed for IMAP-Lo are exclusively H and O, consistent with a nearly pure mono-layer of hydroxyl group (particularly water group) compounds  on the conversion surface. The higher energy of sputtered products likely reflects the lower energy required to liberate loosely bound hydroxyl molecules.  
\end{itemize}

Note that the primary ISN He enters the instrument at 1 au with an energy of $\sim$129 eV. There is some variability in this energy depending on the exact observation time and the PPM angle. Because the sputtered products from the conversion surface have energies lower than the incident atom, we observe the peak sputtering rates (and best statistics) for ISN He at ESA step 3 with a center energy of $\sim$ 56 eV. For the analysis of ISN parameters used in this paper, we focus on observations predominantly from ESA step 3.

\subsection{Operations}

Table 1 shows the PPM angles (the angle between the S/C Z axis and boresight) utilized throughout the mission phases studied in this paper. 
The Science Demonstration period from 2025, DOY 342 -- 2026, DOY 001 was when the year 1 observation plan was exercised. 
There were several instrument outages that led to data gaps. 
The source of these outages was resolved through software changes, and they are are never expected to recur.

The operation of IMAP-Lo during its first year of operations involves pivoting the instrument at predominantly three angles (see Figure \ref{fig:pivot}): 75$^\circ$, 90$^\circ$, and 105$^\circ$. In addition, the spin-axis is offset by 4$^\circ$, on average, from the Sun. This implies that the elongation angles used by IMAP-Lo are $79^\circ$, $94^\circ$, and 109$^\circ$. The stepping of IMAP-Lo pivot platform positions moves each day from 90$^\circ$ on day $N$,  105$^\circ$ on day $N+1$, 90$^\circ$ on day $N+2$, 75$^\circ$ on day $N+3$, and then back to 90$^\circ$ on day $N+4$. The pattern then repeats successively. 

\begin{table}[ht]
\caption{Pivot Platform Positions }
\centering
\begin{tabular}{llll}
\hline
Year & DOY & Pivot angle & Comment \\
\hline
2025 & 312--341 & $90^\circ$ & No PPM motion \\
2025 & 342--364 & $90^\circ$ & even DOY \\
2025 & 343--363 & $105^\circ$ & Every 4th odd DOY (343, 347, \ldots, 363) \\
2025 & 345--365 & $75^\circ$ & Every 4th odd DOY (345, 349, \ldots, 365) \\
2026 & 001 & $90^\circ$ & No PPM motion\\
2026 & 002--016 & $90^\circ$ & E-Box housekeeping outage, data gap \\
2026 & 017--031 & $90^\circ$ & No PPM motion \\
2026 & 032--062 & $90^\circ$ & even DOY \\
2026 & 033--061 & $75^\circ$ & Every 4th odd DOY (033, 037, \ldots, 061) \\
2026 & 035--063 & $105^\circ$ & Every 4th odd DOY (035, 039, \ldots, 063) \\
2026 & 064--078 & $75^\circ$ & E-Box housekeeping outage, data gap \\
2026 & 078--136 & $90^\circ$ & even DOY \\
2026 & 079--135 & $105^\circ$ & Every 4th odd DOY (079, 083, \ldots, 135) \\
2026 & 081--133 & $75^\circ$ & Every 4th odd DOY (081, 085, \ldots, 133) \\
\hline
\end{tabular}
\end{table}

\subsection{Data Products used for He Analysis}

We turn to the data products used in this paper. IMAP-Lo observes both energetic neutral atoms (ENAs) and interstellar neutral (ISN) atoms by measuring H$^-$ and O$^-$ counts, as detailed by \cite{Schwadron:2025}. These two species are distinguished through their time-of-flight (TOF) distributions: at a given ESA step, H$^-$ atoms have higher speeds and therefore shorter TOFs than O$^-$ atoms. ISN He entering the sensor produces predominantly H$^-$ ions through surface sputtering, making H$^-$ rate data the primary observable for deciphering He flow properties. This approach parallels the analysis techniques previously applied to IBEX-Lo data \cite[]{schwadron:2022}.

This paper utilizes two essential IMAP-Lo data products: direct events (DE) and histogram data. DE data represent individual particle counts measured by IMAP-Lo, with each event recording the detection time, pointing direction (boresight), the four measured TOFs \cite[]{Schwadron:2025}, and additional contextual information. The pointing direction for each DE is determined from three components: the spacecraft spin axis direction, the instrument's PPM angle, and the spin phase angle. The spin phase angle is measured event-by-event relative to a spin-pulse timing marker determined by the IMAP-Lo flight software.

Histogram data provide complementary information by binning events into 60 angular sectors, each 6$^\circ$ wide in spin phase. These histograms are further categorized by different TOF combinations, allowing separation of various ion species. For this analysis, we primarily utilize H triple histogram data, which includes only those counts that register all valid times-of-flight and fall within the TOF ranges associated with H$^-$ atoms successfully entering and traversing the complete IMAP-Lo TOF system.



\section{Crossing the 4D ISN Parameter Tubes}

The ISN parameter tubes developed from IBEX-Lo observations involve the interstellar speed given by the ecliptic longitude and latitude, the direction, and interstellar temperature. There is a precise mathematical definition of the parameter tubes that we summarize here. Conceptually, it is helpful to consider a  physical analog. If we are viewing a distant object, a parallax measurement can be made by viewing the object at multiple locations, where these locations are separated along a baseline roughly perpendicular to the line-of-sight. If we view the flow at only one location, we must admit a range of possible characteristic solutions:  a faster flow is less  deflected by the Sun's gravity, causing the observed direction to be closer to that in the upstream region far from the Sun. Conversely, in a slower flow the atoms experience a larger deflection from the upstream direction. This illustrates that the observed speed of interstellar atoms, and their observed direction are linked, causing a significant degeneracy: with only a measurement of the atoms speed and direction at one location, it becomes difficult to establish an absolute pointing direction, or an absolute speed for the upstream flow. 

\begin{figure}
\centering
\includegraphics[width=0.6\columnwidth]{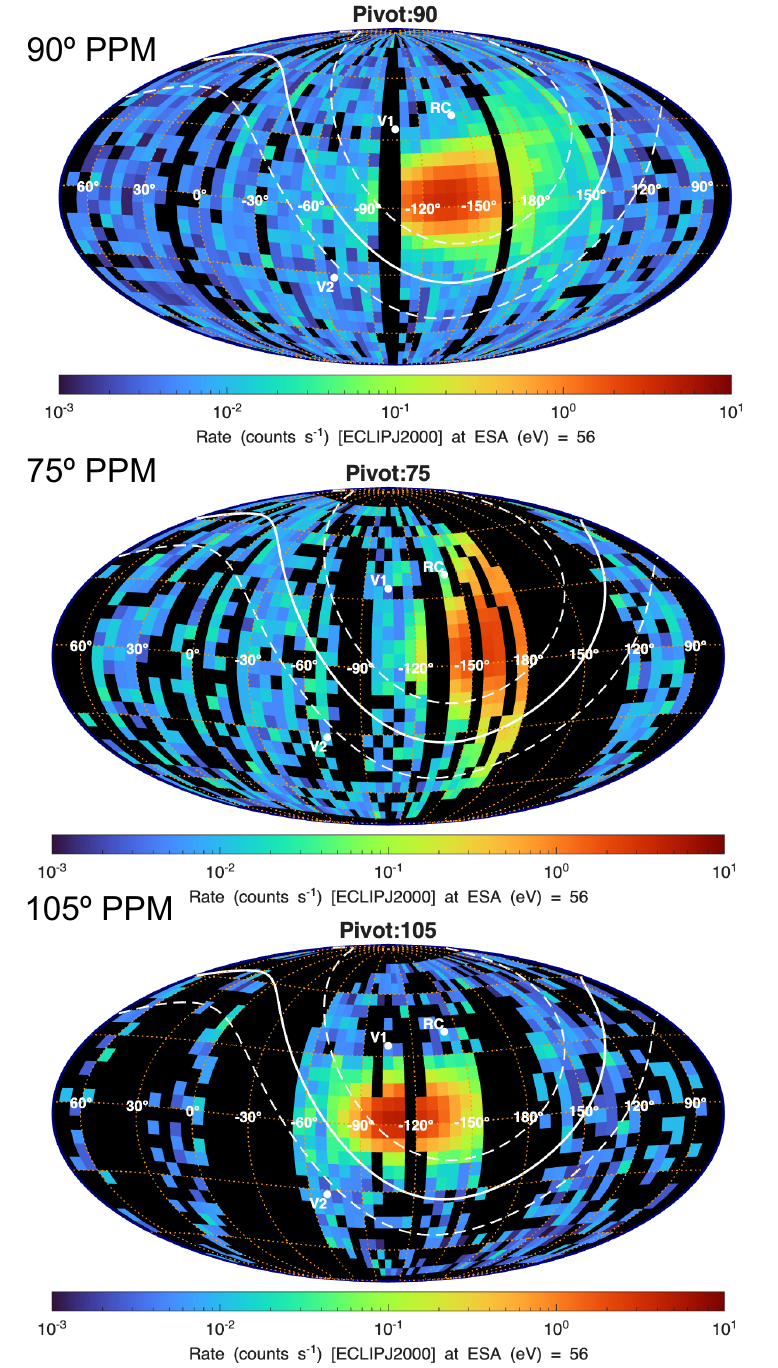} 
\caption{
Count rate maps of sputtered H from incident ISN He at 90$^\circ$ (top), 75$^\circ$ (middle), and 105$^\circ$ (bottom) PPM angle. All rate maps are shown at ESA step 3 with a central energy of 56 eV.  The peaks within these maps are associated with interstellar flow, and the precise location of these peaks is ordered by the elongation angle (the angle between the solar direction and the boresight direction). The elongation angles are, on average 4$^\circ$ larger than the PPM angles since IMAP-Lo directs the spin-angle roughly 4$^\circ$ off from the solar direction to reduce  backgrounds caused by the Sun's UV and visible light. 
 }
\label{fig:maps}
\end{figure}

On IMAP-Lo, by being able to observe the ISN atoms at multiple positions throughout the year, and with multiple viewing orientations, we can infer directly how much deflection there is in the interstellar flow. This therefore provides an absolute direction, and an absolute speed of the interstellar flow. H maps resulting predominantly from ISN He sputtering on the conversion surface are shown for ESA step 3 in Figure \ref{fig:maps}. Note the significant differences in the location of the peak as a function of pivot angle. This is a result of viewing the ISN flow at different elongation angles.

\cite{schwadron:2022} developed a general methodology to  derive the VLISM flow parameters from IBEX-Lo data. This methodology is applied here to IMAP-Lo data. In Appendix B, a series of analytical expressions are used to evaluate the kinematic solutions associated with the trajectories of neutral atoms. The peak physical longitude $\lambda_0$ and peak latitude $\psi^\prime_0$ \cite[the variable definitions are taken from ][]{schwadron:2022} in the observer frame are derived from  the observed properties of inflowing neutral atoms. The kinematic solutions of the observed peaks in longitude and latitude depend on the interstellar parameters: the directional ecliptic longitude $\lambda_\mathrm{ism}$, latitude $\beta_\mathrm{ism}$, the speed $V_\mathrm{ism}$ the temperature $T_\mathrm{ism}$, all referenced to an asymptotic distance far from the Sun. The peak locations also depend on the elongation angle $\eta$. While these relations are straightforward to express, they are difficult to invert. We must find the interstellar parameters based on the observed peak longitude and latitude in the PSD in the observer frame. 

\begin{figure}
\centering
\includegraphics[width=0.8\columnwidth]{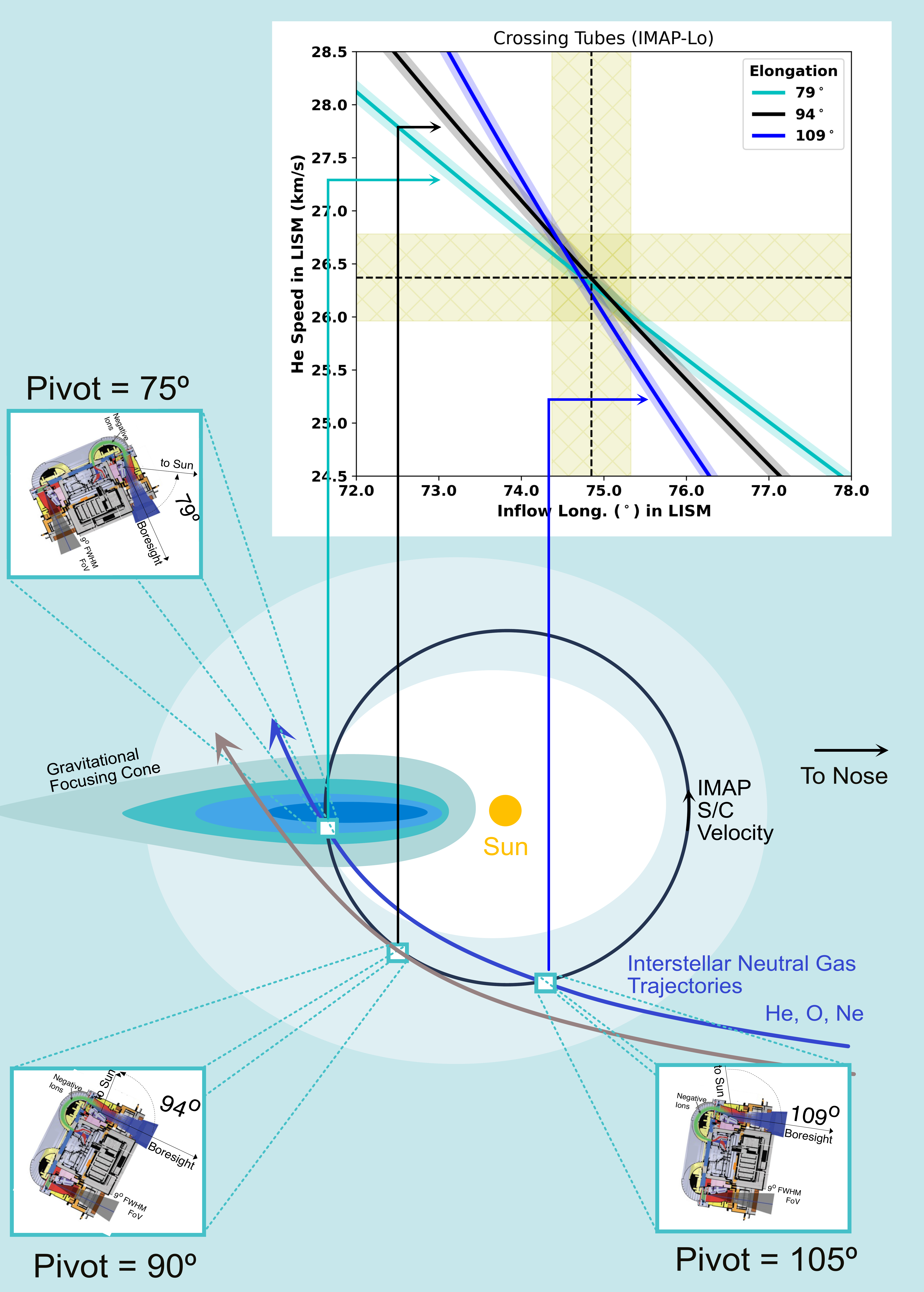} 
\caption{
IMAP-Lo enables determination of ISN parameters using a pivot platform to view the interstellar flow from  multiple orientations. Observations from each elongation angle (the angle between the boresight direction and the direction to the Sun) specify a relationship between the ISN parameters that, in turn,  creates a degeneracy that prevents us from determining a precise and robust set of interstellar parameters. This relationship between interstellar parameters is referred to as the ISN parameter tube, and is shown here with the flow speed as a function the longitude direction of the flow. Each elongation angle is tied to a different parameter tube, and the parameter tubes intersect at the actual interstellar parameters. We show the three primary orientations of the IMAP-Lo instrument during the first year of operations when pivoting enables elongation angles centered on 79$^\circ$, 94$^\circ$, and 109$^\circ$. While representing a hypothetical intersection of the parameter tubes,  the results roughly match those derived in \S 5.
 }
\label{fig:pivot}
\end{figure}

\begin{figure}
\centering
\includegraphics[width=0.8\columnwidth]{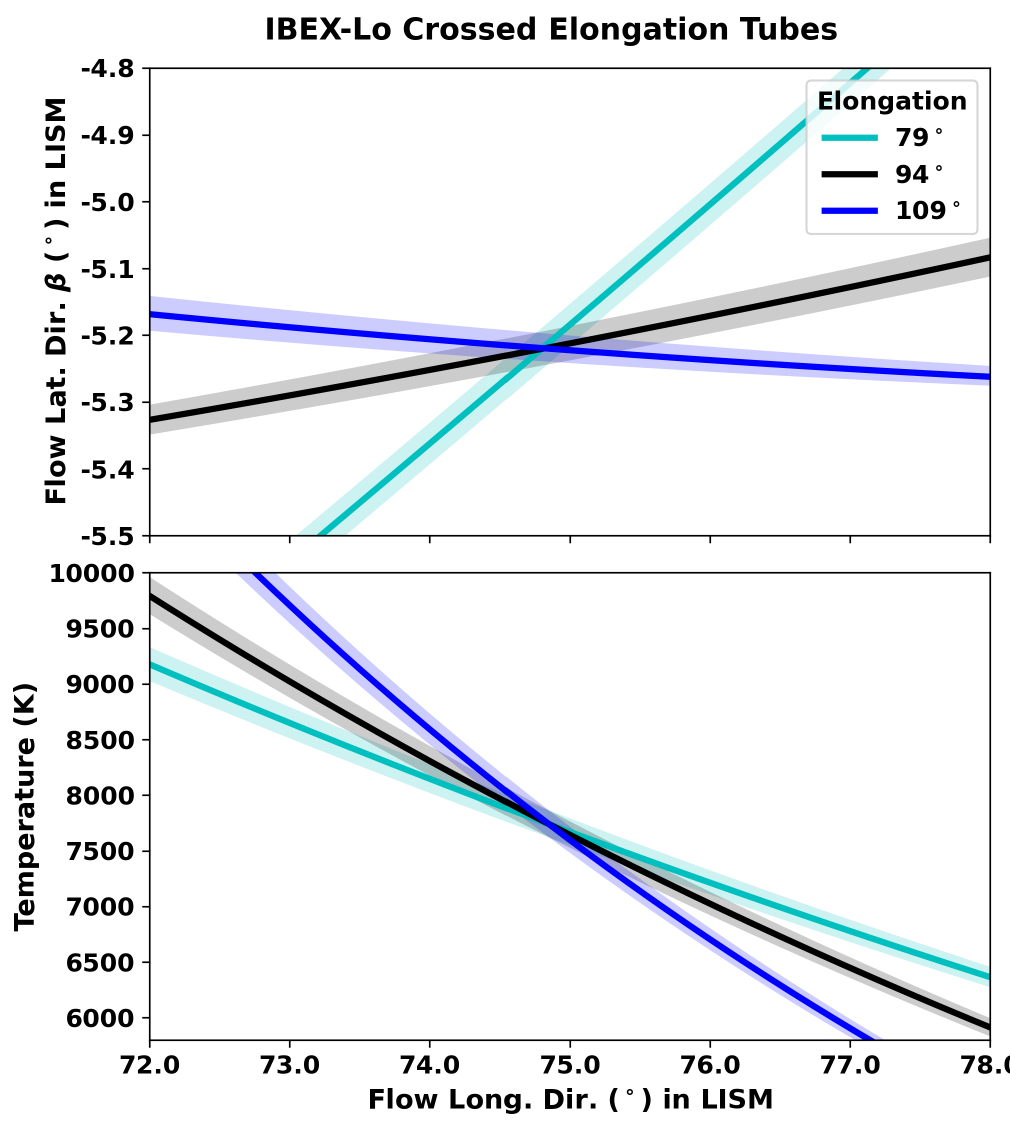} 
\caption{The parameter tube functions of the directional flow latitude (top panel), and the temperature (bottom panel). 
These functions accompany the interstellar speed  solution graphed in Figure \ref{fig:pivot}.   While representing a hypothetical intersection of the parameter tubes, the results roughly match those derived in \S 5.
}
\label{fig:crossingTubes2}
\end{figure}

The approach to the problem used by  \cite{schwadron:2022} is to derive the interstellar longitude direction $\lambda_\mathrm{ism} = \lambda_\mathrm{ism}( V_\mathrm{ism}; \eta, \lambda_0, \psi^\prime_0) $, and the latitude direction $\beta_\mathrm{ism} = \beta_\mathrm{ism}( V_\mathrm{ism}; \eta, \lambda_0, \psi^\prime_0) $ as functions of the interstellar flow speed with fixed parameters for the elongation angle, and the peak longitude and latitude. These relations define the 4D parameter tube that passes through a set of reference interstellar parameters. The  resulting parameter tubes are shown by the  curves and shaded regions in Figures \ref{fig:pivot} and \ref{fig:crossingTubes2}. Fifteen years of observing interstellar neutral atoms with IBEX were critical for establishing for establishing these parameter tube relationships. 

The light and darker blue curves in Figures \ref{fig:pivot} and \ref{fig:crossingTubes2} show the 4D parameter tubes at the additional elongation angles (79$^\circ$ and 109$^\circ$) observed through the first year of IMAP-Lo observations. We have not yet shown the peak longitude and latitude in these distributions, and until those parameters are established, the degenerate parameter tubes cannot be determined. These Figures, though illustrative, point to the key results derived in this paper. In simple terms, ``X marks the spot''',  the crossing points of the parameter tubes define the final solution for the interstellar flow. 

We express the  Gaussian width of the observed distribution \cite[]{lee:2012}:
\begin{eqnarray}
\sigma_{\psi\psi}^2 = \frac{t}{2} \frac{ (v_\mathrm{ism}^2 + 1)^2}{v_\mathrm{ism}^4 [ \sqrt{v_\mathrm{ism}^2 + 2 } +1 ]^2} + \sigma_\mathrm{I}^2
\label{eq:temp}
\end{eqnarray}
where $v_\mathrm{ism} = V_\mathrm{ism} / V_1$, $t = 2 kT_\mathrm{ism} / (m V_1^2)$, $k$ is the Boltzmann constant, and $m$ is the mass of observed He atoms. The quantity $V_1$ is the speed associated with the motion of the Lagrangian L1 point around the Sun, and $\sigma_I = 3.7^\circ$ is associated with a Gaussian-like angular width used to approximate the angular collimation of the incident neutral atoms. (The full-width at half maximum associated with the collimator is 8.71$^\circ$.) Given an observed Gaussian width $\sigma_{\psi\psi}$, equation (\ref{eq:temp}) is readily inverted to derive the interstellar temperature. Note the dependence on the speed $v_\mathrm{ism}$ also introduces a similar parameter degeneracy in the temperature (bottom panel,  Figure \ref{fig:crossingTubes2} ). 

The problem we must solve is more clear: the essential next step is to break the degeneracy in the ISN parameter tube by using observations from multiple elongation angles.  We  find the peaks in the He ISN distributions in longitude, and latitude at each elongation angle observed and then vary solutions along the parameter tube to find the best least-squares fit for the interstellar parameters. 

\section{Longitude and Spin phase Distributions }
\label{sec:lambda}

Primary ISN atoms entering IMAP-Lo sputter atoms predominantly from ESA step 4 and down through the lower ESA steps \cite[see Figure 4 and Figure 2 of ]{schwadron:2022}.
The ISN count rate is typically largest near the peak of the distribution in ESA step 3. Therefore, this ESA step is used 
to determine the parameter tubes at each observed elongation angle.  

\begin{figure*}[ht!]
\centering
\includegraphics[width=0.6\columnwidth]{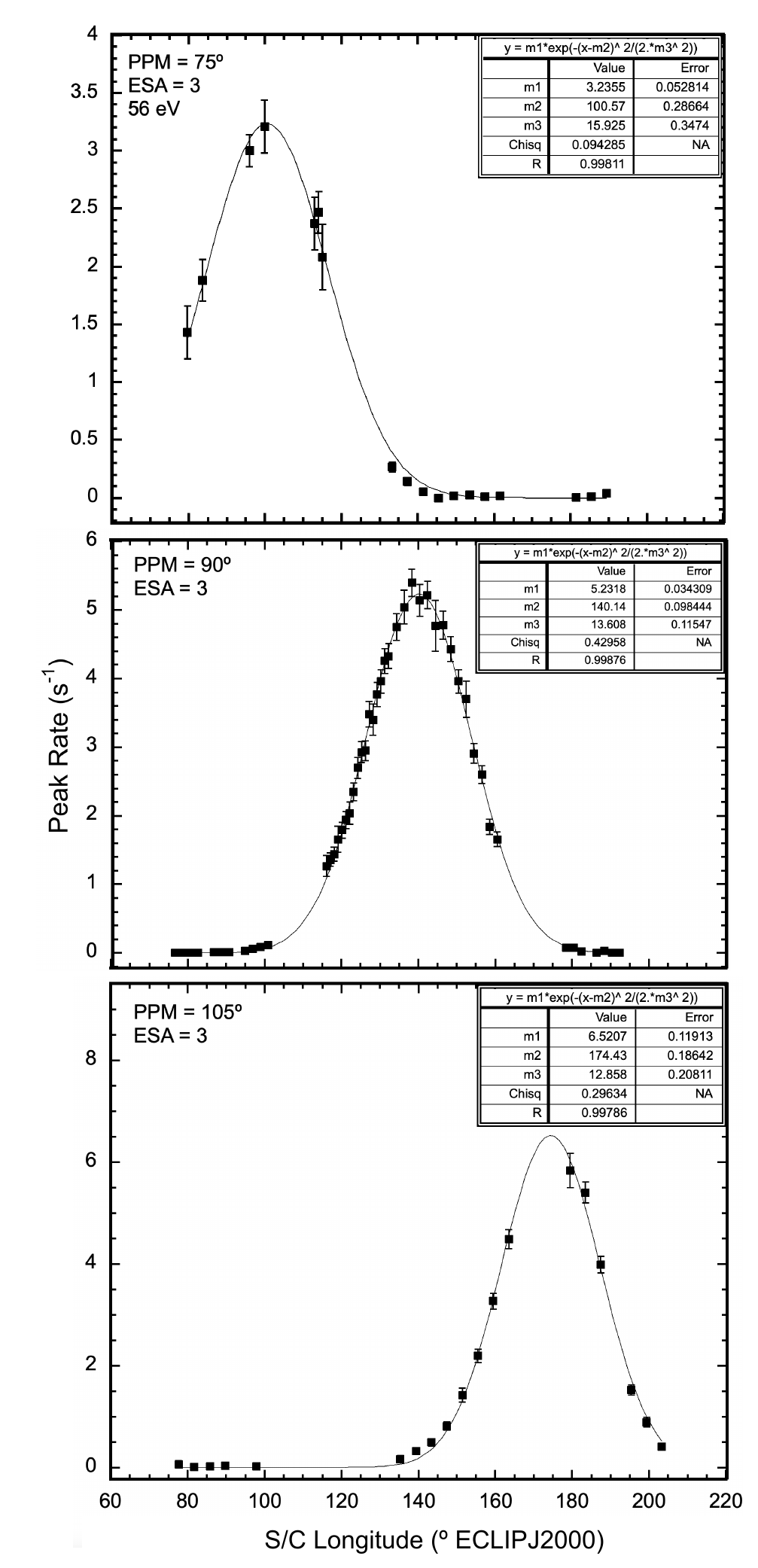}
\caption{The distributions in ecliptic longitude of H count rates at ESA step 3 (56 eV) at PPM angles of 75$^\circ$, 90$^\circ$, and 105$^\circ$. 
The distributions were fit to Gaussian distributions, and the parameters and uncertainties are included in the upper right legends. 
 }.
\label{fig:long_fits}
\end{figure*}

The first step in the determination of the ISN parameters at each observed elongation angle is to find the peak longitude. 
We use H counts observed by IMAP-Lo in histograms in 6$^\circ$ bins. The count rates in each bin are adjusted for the following \cite[]{schwadron:2022}:
\begin{enumerate}
\item contributions of secondary neutral He to the observed count rates;
\item conversion from energy flux density, which is proportional to count rate, to phase-space density (PSD);
\item ionization loss of the ISN He atoms along their trajectories toward the Sun (and IMAP);
\item the sputtering efficiency.
\end{enumerate}
The detailed correction factors  are discussed in Appendix A. Note that item 4 was not applied for IBEX-Lo \cite[]{schwadron:2022}, since the sputtering efficiency was relatively flat at ESA step 3 and below. In contrast, on IMAP-Lo we observe significant changes in the relative response on IMAP-Lo as a function of the ratio of the respective ESA center energy to the incident ENA energy. A correction factor is used to compensate for this effect.  The resulting distributions, after corrections have been applied,  are shown in Figure \ref{fig:long_fits} along with the Gaussian fits. 

\begin{figure*}[ht!]
\centering
\includegraphics[width=0.6\columnwidth]{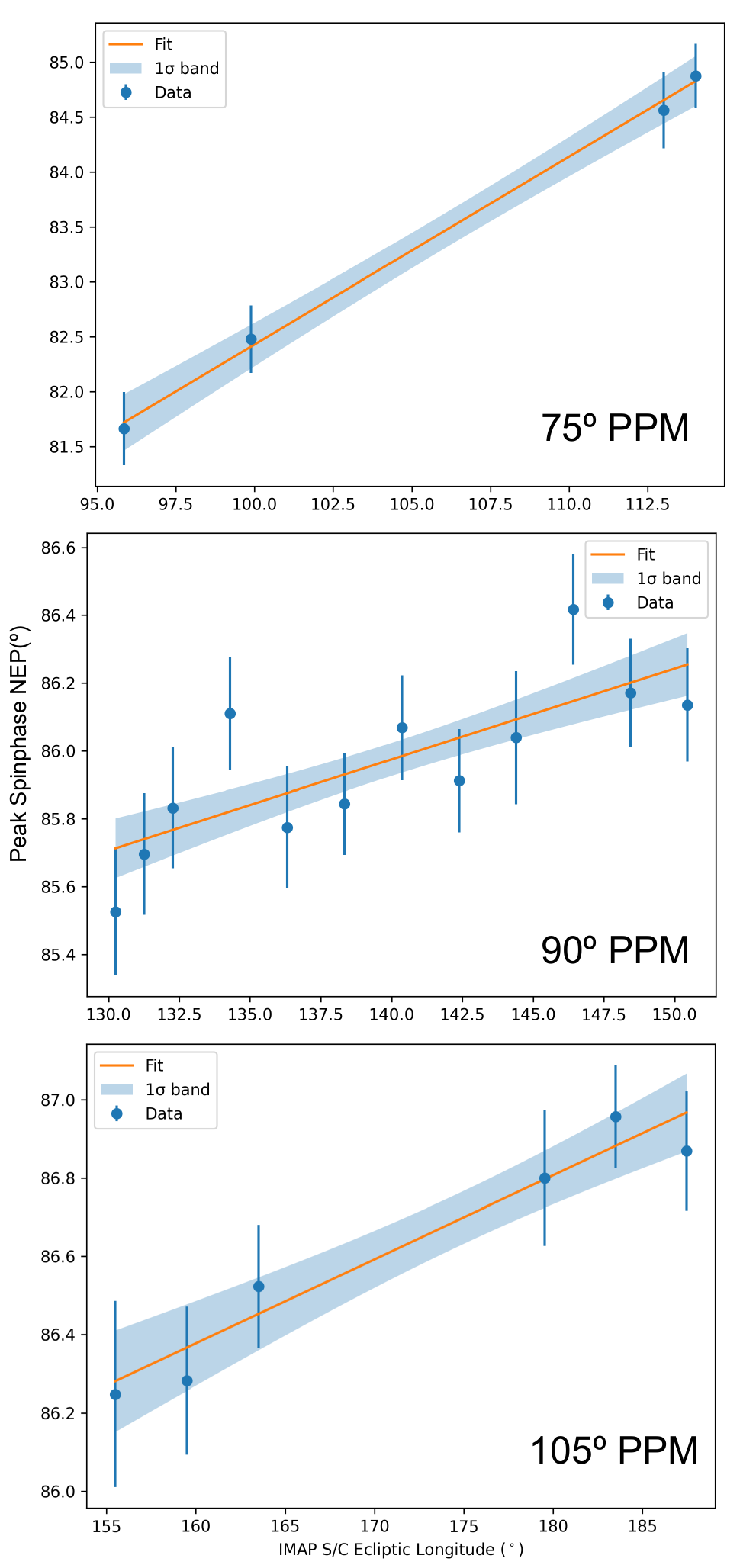}
\caption{The observed H spin phase distributions are fit to a Gaussian at ESA step 3 (56 eV) and PPM angles of 75$^\circ$, 90$^\circ$, and 105$^\circ$. 
We show the Gaussian peak angle  from the North Ecliptic Pole (NEP) as a function of the S/C longitude. The distributions vary linearly, and we 
use the linear fit to determine the peak at the maximum in the distribution. 
 }
\label{fig:lat_fits}
\end{figure*}

The observed H spin phase distributions are also fit to Gaussians. The Gaussian peaks latitudes of these 
distributions follow roughly linear functions of S/C longitude (Figure \ref{fig:lat_fits}). 
We use the linear fits, results of which are listed in Table \ref{tab:fits}, to determine the peak NEP angle at the maximum in the 
distribution. The final estimates of the spin phase peaks in Table \ref{tab:fits} are determined based on the observed maxima in the interstellar flow at each observed  elongation angle. 

\begin{figure*}[ht!]
\centering
\includegraphics[width=0.6\columnwidth]{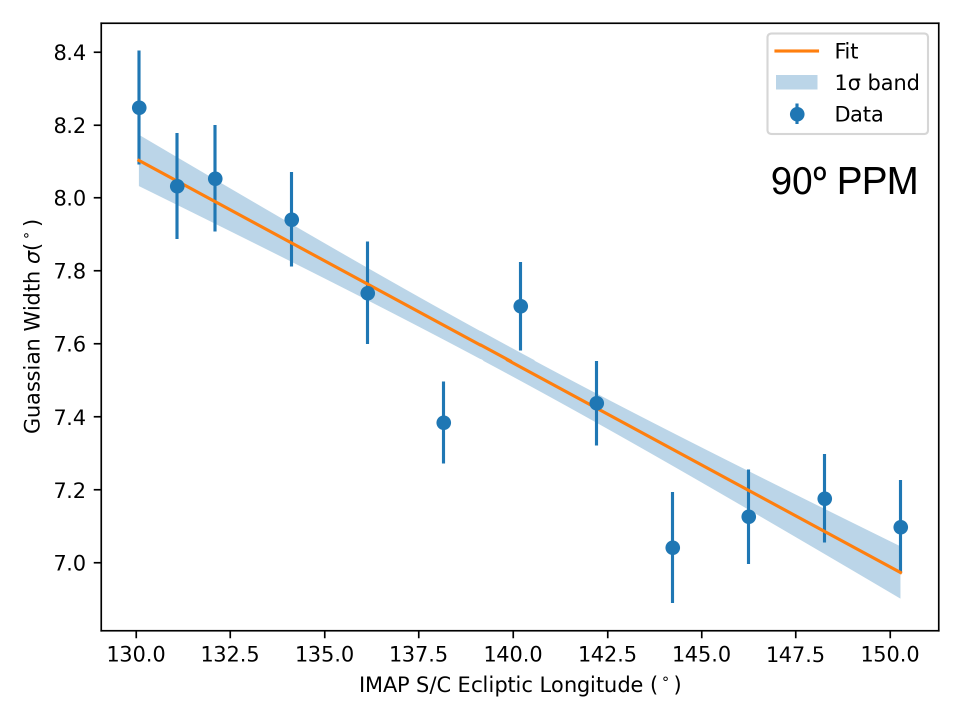}
\caption{The observed Gaussian width of the H spin phase  distribution at ESA step 3 (56 eV) and PPM angles of 90$^\circ$. Gaussian fits to the latitude spin phase distributions yield widths that exhibit a linear trend with S/C longitude. To determine the width at the distribution peak, this linear relationship is interpolated to $140.14^\circ \pm 0.10^\circ$ in S/C longitude (the known peak location), yielding an estimated Gaussian width of $7.55^\circ \pm 0.05^\circ$. 
 }
\label{fig:lat_width}
\end{figure*}

IMAP-Lo also observes the Gaussian widths of the spin phase distributions (Figure \ref{fig:lat_width}). These widths are best determined using the 90$^\circ$ pivot angle, which enables viewing along a great circle aligned with lines of constant longitude (at other elongation angles, the viewing geometry is skewed relative to constant longitude, reducing the apparent angular width of the distribution at pivot angle 105$^\circ$ and increasing it at pivot 75$^\circ$). Gaussian fits to the latitude spin phase distributions yield widths that exhibit a linear trend with S/C longitude. To determine the width at the distribution peak, this linear relationship is interpolated to $140.14^\circ \pm 0.10^\circ$ in S/C longitude (the known peak location), yielding an estimated Gaussian width of $7.55^\circ \pm 0.05^\circ$. 

\begin{table}[ht]
\caption{Linear fits (columns 3,4) to the spin phase peaks and Gaussian width of the distribution shown in Figures 9 and 10.  Derived NEP peaks, and width of the spin phase distributions are listed in columns 5 and 6. \label{tab:fits}}
\centering
\begin{tabular}{llllll}
\hline
Quantity & PPM ($^\circ$) & Slope ($^\circ/$SC long) &  Value ($^\circ$) at 0$^\circ$ SC long & S/C long $^\circ$ at max. &  NEP peak ($^\circ$) at max.  \\
\hline
spin phase Peak & $75$ & $0.171\pm 0.020$ & $65.3 \pm 2.1$ & $100.57 \pm 0.29$ & $82.58\pm 0.35^a$  \\
spin phase Peak & $90$ & $0.027\pm 0.008$ & $82.2 \pm 1.1$ & $140.14 \pm 0.10$ & $86.13\pm 0.34^a$ \\
spin phase Peak & $105$ & $0.022\pm 0.006$ & $82.9 \pm 1.0$ & $174.43\pm 0.19$ & $86.84\pm 0.34^a$ \\
Gaussian width & $90$ & $-0.056\pm0.006$ & $15.4\pm 0.8$ & $140.14 \pm 0.10$ &  $7.55 \pm 0.05$ \\
\hline
\end{tabular}
\medskip
\small
\newline
\textit{$^a$: The derived values for the spin phase peaks shown in the Table have been increased by $0.15^\circ \pm 0.33$ in NEP angle based on a combination of known shifts indicated in Table 8. }
\end{table}

We have added a $0.15^\circ \pm 0.33^\circ$ shift in NEP angle to the peaks derived from H histogram data in the final column of Table \ref{tab:fits}. This shift takes into account an electronic latching found in the flight software that slightly shifts the bin boundaries of histogram data, an electronic lag in S/C pulse-per-second (PPS)  data that determine the spinpulse, and a mounting offset in the star sensor boresight relative to the IMAP-Lo boresight, as detailed in Appendix B (summarized in Table 8). The  electronic lag in S/C PPS data was inferred from the star sensor: the spin phase directions of observed star sensor peaks  were compared to absolute reference directions of known stars from the Tycho 2 catalog \cite[]{Hog:2000}, and the shifts were used to establish the observatory phase lag. 

\section{Interstellar Flow Parameters Determined from ISN He Observations: Speed, Direction, and Temperature}

\begin{figure*}[ht!]
\centering
\includegraphics[width=0.9\columnwidth]{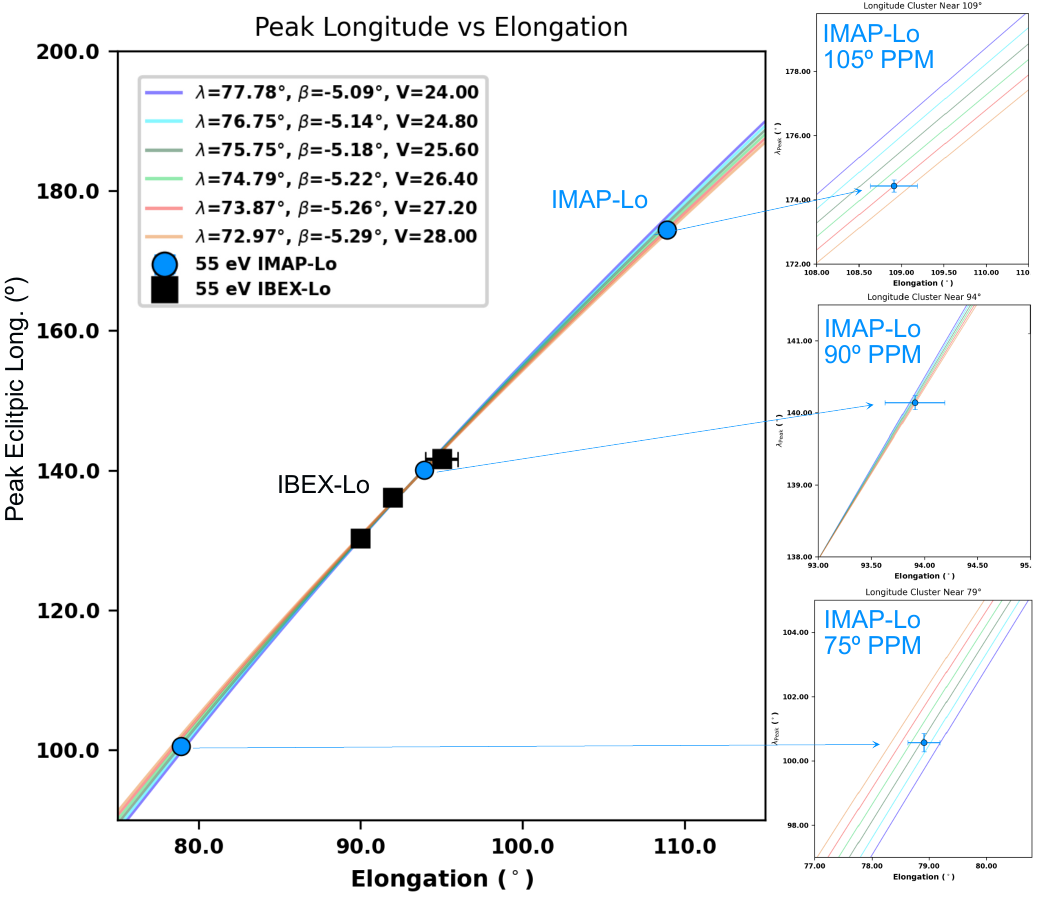}
\caption{The IMAP-Lo (blue round points) and IBEX-Lo (black square points) peak longitudes as a function elongation angle. 
The predictions from interstellar parameters along the parameter tube  are shown by the colored curves.  The IBEX-Lo data points correspond to the peak longitudes derived by \cite{schwadron:2022}. 
 }.
\label{fig:long_tubes}
\end{figure*}

\begin{figure*}[ht!]
\centering
\includegraphics[width=0.9\columnwidth]{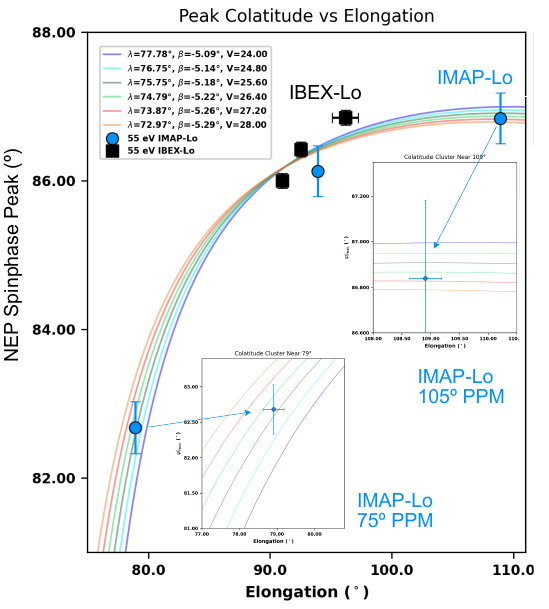}
\caption{The IMAP-Lo (blue round points) and IBEX-Lo (black square points) peak NEP angles as a function elongation angle. 
The predictions from interstellar parameters along the parameter tube  are shown by the colored curves.  The IBEX-Lo data points correspond to the peak latitudes derived by \cite{schwadron:2022}. 
 }
\label{fig:nep_tubes}
\end{figure*}

The procedure for deriving interstellar parameters involves a $\chi^2_r$ search across the parameter tube. Since the parameter tube determines the flow speed and flow latitude direction as a function of the longitude direction, the search is done in only one dimension. Fundamentally, the IMAP-Lo data points are the observed longitude peaks and latitude peaks indicated in Table \ref{tab:fits}.  We show the peak longitude data (Figure \ref{fig:long_tubes}) and the peak NEP angles (Figure \ref{fig:nep_tubes} as a function of elongation angle. The elongation angle is roughly the PPM angle with 4$^\circ$ added due to the average offset of the spin-axis with respect to the solar direction.  This offset significantly reduces the background from aberated solar wind. We also incorporate the mounting offset  (Appendix C) $-0.095^\circ \pm 0.035^\circ$ with respect to the nominal PPM angle. For example, for a nominal 90$^\circ$ PPM angle, the elongation angle is $90^\circ + (4^\circ \pm 0.289^\circ) - (0.095^\circ \pm 0.035^\circ) = 93.91^\circ \pm 0.29^\circ$.  The  $\chi^2_r$ search includes only the IMAP-Lo data points since the IBEX-Lo data were used to derive the parameter tube. 

The predictions from different parameter sets converge near 90$^\circ$ elongation angle, making measurements at PPM angles of 75$^\circ$ and 105$^\circ$ particularly sensitive for constraining the interstellar parameters. We perform a weighted $\chi^2_r$ fit where each data point is weighted according to its observational uncertainty.

\begin{figure*}[ht!]
\centering
\includegraphics[width=0.6\columnwidth]{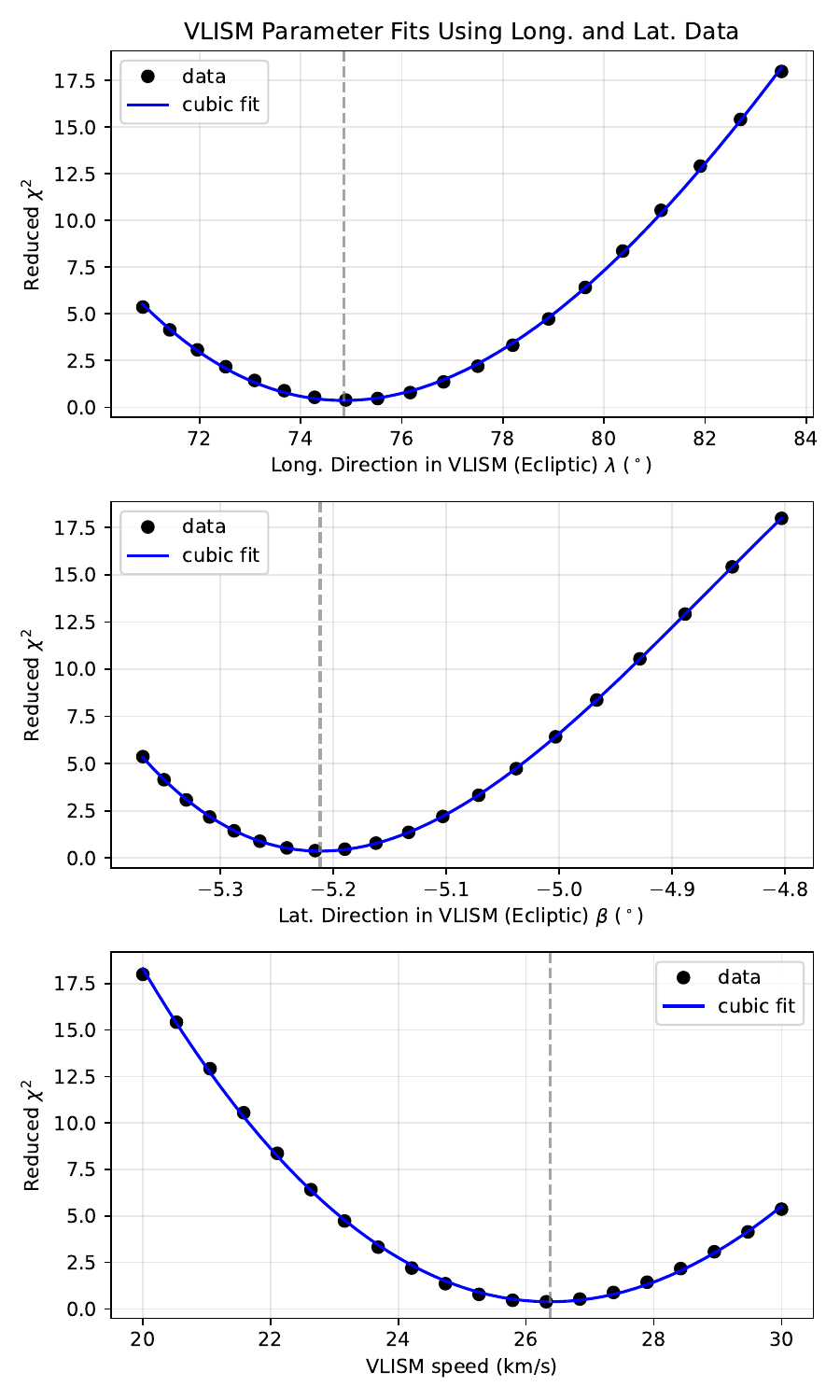}
\caption{The reduced $chi^2_r$ fit using IMAP-Lo peak longitudes and peak NEP angles. Vertical dashed lines show the derived parameters. 
 }.
\label{fig:chisquare}
\end{figure*}

Figure \ref{fig:chisquare} shows the resulting reduced $\chi^2_r$ surface. We derive two sets of interstellar parameters: Table \ref{tab:chi_long} presents results from fitting only the IMAP-Lo longitude peaks, while Table \ref{tab:chi_all} includes all available data.

We calculate uncertainties following the methodology of \cite{Schwadron:2016}, which distinguishes between statistical fit uncertainties ($\sigma_\mathrm{stat}$) and propagation uncertainties ($\sigma_\mathrm{prop}$). For each parameter, we fit the reduced $\chi^2_r$ profile to a polynomial,
\begin{equation}
\chi^2_r = C_0 + C_2 x^2 + C_3 x^3
\end{equation}
where $x$ represents the deviation from the best-fit parameter value. The statistical uncertainty quantifies how well the data constrains the parameter and is given by 
$ \sigma_\mathrm{stat}^2 = \frac{C_0}{\nu C_2}$
where $\nu$ is the number of degrees of freedom in the fit. The propagation uncertainty accounts for the sensitivity of $\chi^2_r$ to parameter variations,
$
\sigma_\mathrm{prop}^2 = \frac{1}{\nu C_2}.
$
The total uncertainty for each parameter is then,
$
\sigma_\mathrm{total} = \sqrt{\sigma_\mathrm{stat}^2 + \sigma_\mathrm{prop}^2}.
$


\begin{deluxetable*}{rlllllll}
\tablewidth{0pt}
\tablecaption{Reduced $\chi^2_r$ cubic fit and derived uncertainties using only IMAP-Lo longitude peaks \label{tab:chi_long}}
\tablehead{
\colhead{Parameter} & \colhead{Param Min} & \colhead{$C_0$} & \colhead{$C_2$} & \colhead{$C_3$} & \colhead{$\sigma_\mathrm{stat}$} & \colhead{$\sigma_\mathrm{prop}$} & $\sigma_\mathrm{tot} $ 
}
\startdata
speed (km s$^{-1}$) & 26.44 &  0.522 & 0.627 & 2.0e-3 & 0.65	& 0.89 & 1.10  \\
long dir ($^\circ$) & 74.82    &  0.522 & 0.456 & -0.014 & 0.75	& 1.04 & 1.29 \\
lat dir ($^\circ$) & -5.212     &  0.522 & 263.0 & -277.7  & 0.032	& 0.044 & 0.055 \\
\enddata
\end{deluxetable*}

\begin{deluxetable*}{rlllllll}
\tablewidth{0pt}
\tablecaption{Reduced $\chi^2_r$ cubic fit and derived uncertainties using only IMAP-Lo longitude and peak NEP angles  \label{tab:chi_all}}
\tablehead{
\colhead{Parameter} & \colhead{Param Min} & \colhead{$C_0$} & \colhead{$C_2$} & \colhead{$C_3$} & \colhead{$\sigma_\mathrm{stat}$} & \colhead{$\sigma_\mathrm{prop}$} & $\sigma_\mathrm{tot} $ 
}
\startdata
speed (km s$^{-1}$) & 26.37 &  0.387 & 0.409 & -0.005 & 0.44	& 0.70 & 0.82  \\
long dir ($^\circ$) & 74.85    &  0.387   & 0.298 & -0.007 & 0.49	& 0.82 & 0.96 \\
lat dir ($^\circ$) & -5.212     &  0.387 & 239.01 & -230.0  & 0.020	& 0.029 & 0.035 \\
\enddata
\end{deluxetable*}

The  Gaussian width of the He distribution was found to be $7.55^\circ \pm 0.05^\circ $. We use Equation (1) and the interstellar speed of $26.37 \pm 0.82$ km s$^{-1}$ to derive the interstellar temperature $T_\mathrm{ISM} = 7740^{+770}_{-730}$ K. A study is currently underway that models the detailed transmission of the ISN through the IMAP-Lo collimator and through the instrument \cite[]{Rahmanifard:2026}. This model was adapted from the IBEX-Lo response model \cite[]{Rahmanifard:2023} and provides a more accurate estimate of the interstellar temperature. 

\section{Discussion}

IBEX observations over more than 15 years were used to derive a set of ISN parameters that had an inherent degeneracy \cite[]{schwadron:2022}. The parameters were compared to previous determinations performed over three decades. 
The degeneracy arises because observations from IBEX are made over a specific region of space where neutral atoms enter the IBEX-Lo instrument at or near 90$^\circ$ elongation angle (or near the perihelion in the He atoms' kinematic trajectories under the influence of solar gravity). This detection location also relies on the motion of the IBEX spacecraft to be in the opposite direction from the neutral atoms, therefore increasing the neutral atom energy to a maximum value at the perihelion in the atom's trajectory. The degenerate parameter set measured by IBEX is known as the IBEX 4D parameter tube \cite[]{lee:2012, moebius:2012, mccomas:2012}.  Table \ref{tab:interstellar} summarizes the  multi-decade assessment of interstellar parameters by IBEX, Ulysses, EUVE, Prognoz 6, ACE, and STEREO, in addition to the IMAP-derived parameters.  Here, we break the degeneracy and thereby reduce the uncertainties in the measured ISN parameters by an average of about a factor of 2 to 3 (see Table 5).

 \begin{deluxetable*}{llllll}
\tablewidth{0pt}
\tablecaption{Interstellar Conditions derived by IMAP, IBEX, Ulysses, EUVE, Prognoz 6, ACE, and STEREO \label{tab:interstellar}}
\tablehead{
\colhead{Method/instrument} & \colhead{$V_\mathrm{ISN \infty}$ (km s$^{-1}$) } & \colhead{$\lambda_\mathrm{ISN\infty}$ ($^\circ$)} & \colhead{$\beta_\mathrm{ISN \infty}$} & \colhead{$T_\mathrm{ISN\infty}$ (K) } & \colhead{Reference}  
}
\startdata
Neutral Gas/Ulysses-GAS & $26.3 \pm 0.4$ & $75.4 \pm 0.5$ & $-5.2 \pm 0.2$ & $6300 \pm 340$ & \cite{witte:2004b} \\
Neutral Gas/Ulysses-GAS & $26.08 \pm 0.21$  &  $75.54 \pm 0.19$ & $-5.44\pm 0.24$ & $7260 \pm 270$ & \cite{Wood:2015}  \\
Neutral He/IBEX & $25.8 \pm 0.4$ & $75.8 \pm 0.5$ & $-5.16 \pm 0.1$ & $7440 \pm 260$ & \cite{bzowski:2015} \\
Neutral He/IBEX & $25.82 \pm 0.33$ & $75.62\pm 0.36$ & $-5.19 \pm 0.06$ & $7673 \pm 225$ & \cite{swaczyna:2018} \\
UV backscatter/EUVE & $24.2 \pm 2$ & $74.7 \pm 0.5$ & $-5.7 \pm 0.5$ & $6500 \pm 2000 $ & \cite{Vallerga:2004} \\
UV backscatter/Prognos 6 &                 & $74.5 \pm 1$      & $-6 \pm 1$              &                       & \cite{lallement:2004a} \\
\hline 
LIC Absorption/Haute-Provence  & $25.7 \pm 1$ &                  &                           & $7000 \pm 1000$ & \cite{lallement:1992a}  \\ 
LIC Absorption/Hubble                   &                      & &  &  $7000 \pm 200$ & \cite{linsky:1993a} \\
\hline
Pickup Ions/ACE-SWICS &                  & $74.74 \pm 0.33$ &                              &                       & \cite{gloeckler:2004a} \\ 
Pickup Ions/STEREO-PLASTIC &           & $75.41 \pm 0.34 $ &                           &                          & \cite{taut:2018a} \\
\hline 
Param Tube Intersect (IBEX) & $25.99^{+1.86}_{-1.76}$ & $75.28^{+2.27}_{-2.21}$ & $-5.2^{+0.093}_{-0.085}$ & $7496^{+1274}_{-1528} $ & \cite{schwadron:2022} \\
Param Tube Intersect (IMAP) & $26.37 \pm 0.82$ & $74.85\pm 0.96 $ & $-5.212\pm 0.035$ & $7740^{+770}_{-730} $ & This study \\
\hline 
VLISM in Solar Frame (IMAP) & $26.8 \pm 0.8$ & $74.85 \pm 0.96$  & $-5.21\pm 0.04 $ & $6640 \pm 760$ & This study \\
\enddata
\end{deluxetable*}

The derived interstellar parameters after correction  (last row in Table 5) for elastic collisions \cite[]{swaczyna:2021, islam:2025} appear to represent an average between the LIC  ($ T = 7500 \pm 1300$ K) and G-Cloud ($T= 5500 \pm 400$) parameters. The LIC, which is further out in the Loop I shell, is observed to be both slower  and hotter, while the G-cloud  is cooler and faster. Not only is the VLISM an intermediate state, but it appears to be consistent with a slowing and heating in the LISM caused by the interaction between these clouds. This intermediate region  was alluded to by \cite{redfield:2008}: ``The in situ 'Helio' measurement is closer to the average LIC and G temperature than either cloud individually'' and the average value is $6500 \pm 680$ K. Furthermore, the temperature derived here for the VLISM $6640\pm 760$ K  is quite close to the value $6306 \pm 390$ K derived by \cite{Moebius:2004} for the VLISM as a weighted mean between the Ulysses/GAS \cite[]{witte:2004b} and EUVE results \cite[]{Vallerga:2004}. Remarkably, these results were not corrected for elastic collisions, but nevertheless, were largely consistent with IMAP results after correction for elastic collisions (Table \ref{tab:interstellar}). This reinforces the view that the LISM is a medium slowing and heating through intercloud interactions, and indicates the local gradient in the LISM speed and temperature. 

\begin{deluxetable*}{lllllll}
\tablewidth{0pt}
\tablecaption{VLISM state and truth tests\label{tab:tests}}
\tablehead{
\colhead{VLISM state} & \colhead{Speed (km/s)} & \colhead{Test} & \colhead{Temp (K)} & \colhead{Test} & \colhead{Truth} & \colhead{References}
}
\startdata
LIC &   $23.84 \pm 0.9$ 				& \textbf{Fail} & $7500 \pm 1300$ &  Pass   & not LIC & \cite{redfield:2008} \\
G-Cloud & $29.6 \pm 1.1$ 			& \textbf{Fail} & $ 5500 \pm 400$ &   Pass & not  G-Cloud & \cite{redfield:2008} \\
\hline
\textbf{Average}  & $26.74 \pm 0.71$  				& \textbf{Pass} & $6500 \pm 680$ &  \textbf{Pass} & \textbf{Intermediate}  &  \cite{redfield:2008} \\
\hline
10 yr change, mixed &  $\sim -0.013$   & -- & $\sim +4$      & --  & Unknown & this work \\ 
10 yr change, compressed &  $\lesssim -0.00027$   & -- & $\lesssim +1$       & -- & Unknown & this work 
\enddata
\end{deluxetable*}

Table \ref{tab:tests} provides a truth table indicating whether the VLISM is in the LIC, the G-Cloud or an in an intermediate state. 
The resolution of the VLISM speed and temperature using the parameter tube intersection enables us to remove significant uncertainties, definitively demonstrating that the VLISM is not consistent with either the LIC or the G-cloud \cite[]{swaczyna:2022clouds}. However, the observations are consistent with an intermediary state between the G-cloud and the LIC \cite[]{redfield:2008}.

Several scenarios are possible for the VLISM having flow properties that are intermediate between those of the G-cloud and the LIC. In the ``mixed state'', charge-exchange allows neutral particles to be formed and move across plasma boundaries. 
The thickness of the boundary can be estimated as $\kappa \; t_\mathrm{loop-I} \sim 0.08$ pc. Here $ t_\mathrm{loop-I}  \sim$ 1 Myr is the lifetime of the L1 superbubble \cite[]{Wolleben:2007}, and $\kappa = \lambda_{cx} v/3$ is the diffusion coefficient associated with charge-exchange (we take 270 au mean free path, $\lambda_{cx}$, and a particle speed $v \sim 13.5$ km s$^{-1}$ close to the sound speed). Given the 0.08 pc thickness of the boundary and the $\sim 18$ km s$^{-1}$ motion of the heliosphere through the LISM in the LSR frame, we would observe a slight speed reduction by $\sim -0.013$ km s$^{-1}$, and temperature change by $\sim +4$ K over a 10 year period. 

Should such a large change have been observed previously? Previously quoted uncertainties in temperature were typically $\sim 250$ K (Table 5, rows 3-4), which is more than 60 times the temperature change predicted for a mixing boundary in the LISM. The observing strategy to determine this gradient relies on significant uncertainty reduction, which can be achieved on IMAP-Lo by extending the range of pivot platform angles. 

We note that the slowing flow in the LISM  may be driving a plasma compression or shock. Based on the 6640 K  temperature in the VLISM, the sound speed associated with ions and electrons is $\sim 13.5$ km s$^{-1}$, and  the Alfv\'en speed in the VLISM is $\sim 32$ km s$^{-1}$ \cite[]{schwadron:2021}. These speeds are both greater than the speed difference between the G-Cloud and LIC, and both the Alfv\'en speed and the fast magnetosonic speed exceed the plasma speed in the LSR frame.  Therefore, if the LISM is in an interaction region, then this structure is likely a plasma compression. 

Assuming an adiabatic pressure, we form the ratio of the compression pressure gradient $\nabla P$ to the inertial flow gradient $\rho \mathbf{u} \cdot \nabla \mathbf{u}$ in the compression. This ratio is about 36\%, meaning that the pressure gradient is not large enough to create the negative speed gradient observed in the LISM. However, based on the 32 km s$^{-1}$ Alfv\'en speed in this region, we find that the magnetic field pressure is roughly 2.7 times the plasma ion and electron pressure. Therefore, taking into account the LISM magnetic field \cite[]{schwadron:2021}, the total plasma pressure is not just sufficient to slow the flow in the LISM, but also very close (to within 1\%) to maintaining a  balance between the total plasma pressure gradient and the inertial flow gradient in the compression region. This lends significant credibility to the compression region scenario. However, the gradient scale in this case likely extends across the entire region between the G-cloud and the LIC over $> 3.9$ pc. As a result, the observed change in the VLISM over a 10 year period would be very small: $\lesssim -0.00027$ km s$^{-1}$ in plasma speed and $\lesssim +1$ K in temperature. 

The compression scenario for the VLISM also has important ramifications for the ordering of the interstellar magnetic field, and its influence of cosmic ray diffusion \cite[]{schwadron:2014}. 
Differentiation between the interstellar magnetic field on local scales ($\sim$1000 au or $\sim$0.005-0.01 pc) and on larger (parsec) scales is expected due to the presence of turbulence 
in the local galactic medium \cite[]{Lee:1976, minter:1996, Chandrasekhar:1952, Jokipii:1969, Armstrong:1995}. Turbulence
disrupts the LISM flow and 
leads to a complex and tangled magnetic field structure. In the interstellar medium, typical coherence scale lengths 
along which the magnetic field appears relatively ordered are typically 1-10 pc \cite[]{minter:1996}. 

The presence of turbulence raises another important question regarding homogeneity within the LISM. To what degree does the difference between the VLISM state and that of the G-cloud and the LIC reflect inhomogeneity within this medium? Answering this question requires assessing not only VLISM change in time, but also the intrinsic differences in interstellar parameters that may exist among different observed species, such as interstellar O \cite[]{Schwadron:2016} and interstellar H \cite[]{rahmanifard:2019}. 

Compression within the LISM provides an 
important physical driver, which like the heliosphere itself, could act build up and deflect the magnetic field in front of the compression (the ``snowplow" effect akin to coronal mass ejecta, but in the interstellar medium). 
Ironically, the ribbon itself is a feature that partially reflects the ordering of the VLISM magnetic field in the presence of flow compression around the 
heliosphere \cite[]{mccomas:2009, funsten:2009, schwadron:2009, schwadron:2013, zirnstein:2016}. If the VLISM is in a compression, this effect would likely act to add 
order the interstellar magnetic field in the VLISM, drive the field direction to be roughly perpendicular to the flow in the LSR frame \cite[]{schwadron:2014}, and possibly increase the coherence scale length. This would, in turn, affect the propagation of cosmic rays prior to entry into the heliosphere. 

\section{Conclusions}

IMAP-Lo has successfully reduced systematic uncertainties by using its pivot platform to observe the interstellar flow from multiple elongation angles (boresight directions relative to the Sun).
We provide the first results that definitively intersect ISN parameter tubes over elongation angles $79^\circ$, $94^\circ$, and $109^\circ$, resulting in precise interstellar neutral parameters and reduced systematic uncertainties  by a factor of 2 to 3 over the previous IBEX results: speed $26.37 \pm 0.82$ km s$^{-1}$, ecliptic longitude direction $74.85^\circ \pm 0.96^\circ$, ecliptic latitude direction $-5.212^\circ \pm 0.035^\circ$, and temperature $7740^{+770}_{-730}$ K. 

The derived flow state of the VLISM is between that of the LIC and G Cloud, but inconsistent with either cloud. This supports the concept that the VLISM within the broader LISM is in a region that mixes material from G-Cloud and the LIC,  shown to be likely by \cite[]{swaczyna:2022clouds}. It is possible that the LISM  between the G-clould and the LIC is a large interaction region, a compression region, consistent with a decelerating flow in the Loop I shell \cite[]{Frisch:2013}.

The question that IMAP is poised to answer is whether over the coming years we can provide a precise enough determination of plasma speed and temperature to measure and thus understand the physics of the intermediate cloud structure in the LISM. Over the scale of many 100's or 1000's of years, the prediction is that the heliosphere will move closer to the state of the LIC where the flow is slower and the temperature is hotter.  This should induce significant change for the global state of the heliosphere, which could have wide ranging effects for the space environment and Earth \cite[]{schwadron:2011b}.

We conclude that following earlier IBEX observations, IMAP has taken another major step in understanding the state of the Local Interstellar Medium, which defines the outer boundaries of our heliosphere.
Specifically, we have at last removed a difficult degeneracy in the determination of interstellar parameters by observing interstellar neutral atoms at three distinct pivot platform viewing angles. 
In the future, IMAP-Lo will observe the interstellar flow across an even broader range of pivot platform angles to determine even more accurate interstellar flow solutions. 
This work positions us  to understand the detailed physics of the complex intermediate cloud region that constitutes the  Local Interstellar Medium. By resolving and understanding this physical region, we will also determine the outer boundary conditions of the heliosphere and how they will evolve in the future.

\begin{acknowledgments}
This paper is dedicated to our dear friend and colleague, Priscilla Frisch, who helped shape our emerging view of the heliosphere within the local interstellar medium.
We are deeply indebted to everyone who helped make the
IMAP  mission possible. We thank all of the
outstanding scientists, engineers, technicians, and administrative
support personnel  across all of the 
institutions that produced and supported the  instruments, the mission, and support its operations and the scientific
analysis of its data. We also thank the scientists, engineers, and technicians who developed, tested, and operated IBEX and IBEX-Lo. The success of this explorer mission has been an enormous benefit to the development of the IMAP mission and the IMAP-Lo instrument.  This work was supported as a part of the IMAP  mission. Peter Wurz, Michela Gargano, and Andre Galli acknowledge the financial support from PRODEX (PEA 4000129386). M.B. and M.A.K. were supported by Polish National Science Center (NCN) grant 2023/51/B/ST9/01921. P.S. was supported by the Polish National Agency for Academic Exchange within the Polish Returns Programme (BPN/PPO/2022/1/00017) and the National Science Centre, Poland (2025/58/E/ST9/00205)
\end{acknowledgments}

\appendix

\section{Correction factors for observed peak rates}

As discussed in \S \ref{sec:lambda}, we correct peak rates for (1) contributions of secondary neutral He to the observed count rates, 
(2) conversion from energy flux density, which is proportional to count rate, to PSD, (3) ionization loss of the ISN He along its trajectory toward the Sun (and IMAP), and (4) the sputtering efficiency. For each of these quantities, we form a correction factor that multiplies the rate to account for the effect. 
For items (1) - (3), we use the Warsaw Test Particle Model (WTPM) \cite[]{bzowski:2012, bzowski:2015, kubiak:2016} to estimate contributions from secondary atoms, the ISN speed, and ionization loss. The correction to PSD involves dividing the rate by a quantity that depends on the inverse of the He atom speed in the S/C frame \cite[see equation A19 from ][]{schwadron:2022}.  For the sputter efficiency, we use the estimated primary He atom speed and the relative efficiency curve (Figure 4) to develop a correction factor.

\begin{figure}
\centering
\includegraphics[width=0.8\columnwidth]{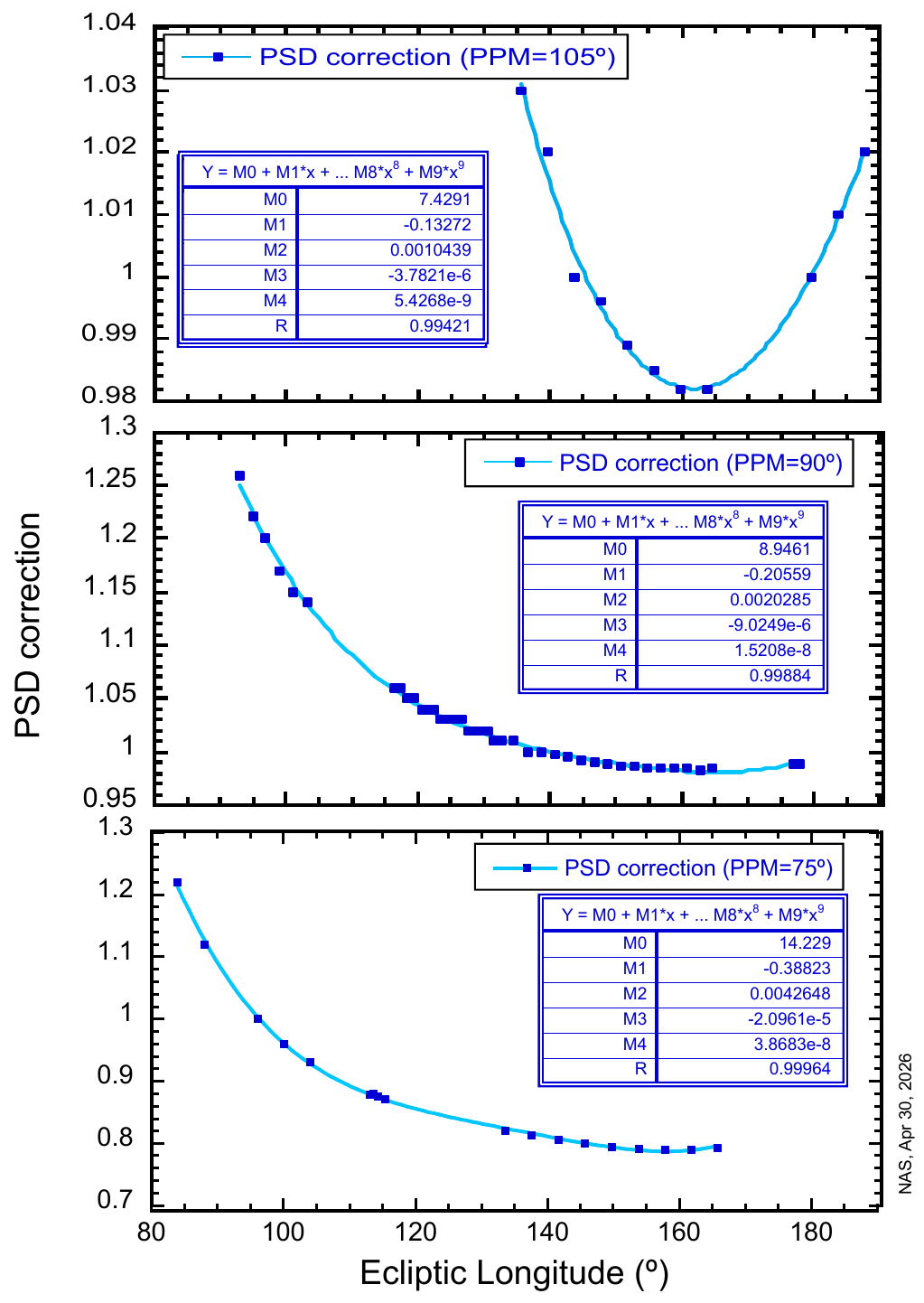} 
\caption{The PSD correction factor  multiplies the observed H count rate to form a quantity proportional to the distribution function. The quantity is found from an estimate of the He at speed at 1 au, and used to estimate the atom speed in the S/C frame. The observed rate is divided by the estimated atom speed in the S/C frame, and then renormalized to form an estimate of the distribution function. A fourth-order polynomial is fit to the correction factor at each pivot angle.
 }
\label{fig:psd_factor}
\end{figure}

Each correction factor is fit to a function of ecliptic longitude. Figure \ref{fig:psd_factor} shows an example of these fits for  the PSD correction factor at each of the utilized pivot angles.  
We use the following general form for polynomial fits for the correction factors:
\begin{eqnarray}
\xi = M_0 + M_1 \lambda + M_2 \lambda^2 + M_3 \lambda^3 + M_4 \lambda^4
\end{eqnarray}
where $\lambda$ is the ecliptic longitude expressed in degrees. 
The coefficients are compiled in Table \ref{tab:fits2}.

\begin{deluxetable*}{lllllll}
\tablewidth{0pt}
\tablecaption{Fit coefficients to correction factors \label{tab:fits2}}
\tablehead{
\colhead{Correction} & \colhead{Pivot Angle ($^\circ$) }& \colhead{$M_0$} & \colhead{$M_1$} & \colhead{$M_2$}  & \colhead{$M_3$} &  \colhead{$M_4$}
}
\startdata
Secondaries & 75  & 0.87241 & -2.9824e-3 & 7.6829e-5 &  -3.7244e-7 & 0  \\
Secondaries & 90 & -42.21 & 1.1625 & -1.1775e-2 & 5.3171e-5 & -9.0282e-8 \\
Secondaries & 105 & 1       & 0 & 0 & 0 & 0 \\
PSD & 75 & 14.229 & -0.38823 & 4.2648e-3 & -2.0961e-5 & 3.8683e-8 \\
PSD & 90 & 8.9461 & -0.20559 & 2.0285e-3 & -9.0249e-6 & 1.5208e-8 \\
PSD & 105 & 7.4291 & -0.13272 & 1.0439e-3 & -3.7821e-6 & 5.4268e-9 \\
Ionization & 75 & 1.7515 & -8.9354e-3 & 1.4313e-5 & 0 & 0 \\
Ionization & 90 & 2.1744 & -1.0734e-2 & 1.6799e-5 & 0 & 0 \\
Ionization & 105 & 2.6894 & -1.333e-2 & 2.0923e-5 & 0 & 0 \\
Efficiency & 75 & 2.7025 & -3.5621e-2 & 2.1462e-4 & 0 & 0 \\
Efficiency & 90 & -6.3543 & 0.14168 & -8.4994e-4 & 1.7032e-6 & 0 \\
Efficiency & 105 & -0.61069 & 3.0789e-2 & -1.5798e-4 & 2.5776e-7 & 0 
\enddata
\end{deluxetable*}

\section{Spin Phase Peak Corrections}

\begin{figure}
\centering
\includegraphics[width=0.8\columnwidth]{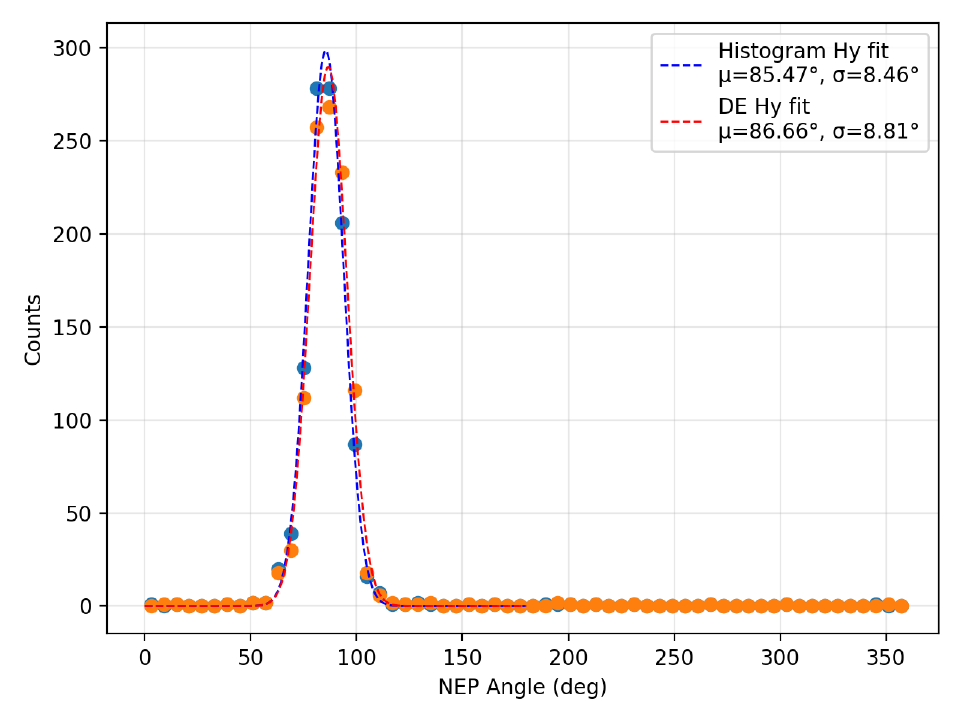} 
\caption{
The observed distribution of H counts in spin phase, obtained separately from Direct Events (orange dots) and Histogram data (blue dots) on January 23, 2026 at ESA step 4 (106 eV). During this period, the spacecraft's ram direction aligned with the ISN flow, producing a sharp peak in ISN Helium counts. The use of ESA step 4, which accepts the majority of primary He atoms, further contributes to the narrow distribution. Note that any broadening of the distribution would reduce the observed offset between the DE and Histogram peaks.}
\label{fig:DE_Hist}
\end{figure}

The analysis of histogram data to derive spin phase shifts requires adjustments to account for three major effects listed in Table \ref{tab:shifts}. Note that we have focused on ESA step 3 which has the highest statistics and provides the best sample of the interstellar flow. The first of the shifts is a $\sim 50$ ms electronic lag in histogram bins (see Figure \ref{fig:DE_Hist}).  The latching effect causes a reduction in the inferred NEP angle since bin $i$ closes 50 ms late. The correction factor of $\sim 1.2^\circ$ listed in Table \ref{tab:shifts} is added the observed NEP angles  to compensate. 

Several additional effects are considered. The star sensor was aligned to the IMAP-Lo boresight using alignment cubes. In the final measurements on the launchpad, a 0.25$^\circ$ mounting error was measured between the star sensor and the IMAP-Lo boresight in the spin phase plane (see Table 9). Lastly, the star sensor data was used as an absolute reference for the observatory based on stars with known celestial positions.   A net lag in the observatory spin phase of $-1.3^\circ \pm 0.3^\circ$ was found. The sum of these three spin phase shifts provides the net offset added to the histogram spin phase peaks of $0.15^\circ \pm 0.33^\circ$. 

A number of actions were taken  to correct these issues. The histogram shift was corrected in IMAP-Lo software version 4.8 uploaded to the instrument on May 22, 2026. The IMAP S/C team has found the electronic lag causing the observatory spin phase offset. The issue has been corrected in S/C flight software on July 8, 2026.

\begin{deluxetable*}{llll}
\tablewidth{0pt}
\tablecaption{Shifts in spin phase \label{tab:shifts}}
\tablehead{
\colhead{Quantity} & \colhead{Cause }& \colhead{Relative To:} & \colhead{spin phase Shift ($^\circ$)} }

\startdata
spin phase peak in Histogram & Electronic latching (ICE$^a$) & Direct Event peak & $1.2  \pm  0.1$   \\
Star Sensor boresight & Mechanical Mounting & IMAP-Lo Boresight & $0.25 \pm 0.05$ \\
Spinpulse & Electronic Lag in S/C Attitude System$^b$ & Star Sensor Boresight & $-1.3 \pm 0.3$ \\
\hline 
Net Shift to Histogram Peak & Sum of Effects & NEP & $0.15 \pm 0.33$  
\enddata
\tablecomments{
\\
$^a$: IMAP-Lo Common Electronics (ICE) \\
$^b$: Pulse-Per-Second (PPS) information from S/C is used by the ICE to predict a spinpulse in each S/C rotation }
\end{deluxetable*}

\section{Mounting Offsets}

The placement of alignment cubes on the star sensor and the IMAP-Lo boresight were used prior to launch to determine the precise alignments of both the star sensor and the IMAP-Lo boresight. Table \ref{tab:alignment} provides the alignment determinations from these pre-launch measurements. 

\begin{deluxetable*}{llllll}
\tablewidth{0pt}
\tablecaption{Alignment of star sensor and IMAP-Lo boresights\label{tab:alignment}}
\tablehead{
\colhead{Boresight} & Nom. PPM angle ($\circ$) &  \colhead{State} &  \colhead{Angle to Y-axis ($^\circ$) }& \colhead{PPM ($^\circ$)} & \colhead{StS to IMAP-Lo ($^\circ$)$^a$}
}

\startdata
IMAP-Lo & 60 & Locked$^b$  & 29.846 & 59.922 &  0.465 \\
Star Sensor & 60 & Locked$^b$ & 30.123 & 60.321 &  \\
IMAP-Lo & 60 & Unlocked$^b$ & 29.848 & 59.947 &   0.467 \\
Star Sensor & 60 & Unlocked$^b$ & 30.126 & 60.346 &  \\
IMAP-Lo & 90 & post 60-to-90$^c$ & 29.888 & 89.880 &   0.467 \\
Star Sensor & 90 & post 60-to-90$^c$ & 30.129 & 90.280 &      \\
IMAP-Lo & 90 & post 179-to-90$^d$ & 29.886 & 89.873 &   0.468 \\
Star Sensor & 90 & post 179-to-90$^d$ & 30.128 & 90.273 &        \\
\hline 
Net Shift IMAP-Lo &  &   & $-0.132 \pm 0.023^e$   &    $-0.095 \pm 0.035^f$ & $0.467 \pm 0.001$ \\
Net Shift Star Sensor$^e$ &  &  & $0.126 \pm 0.002^e$  &  $0.305 \pm 0.035^f$ \\
\enddata
\tablecomments{
\\
$^a$: Angle ($^\circ$) between the Star Sensor and IMAP-Lo boresights \\
$^b$: Locked and Unlocked refer to the Pivot Platform Hold-Down Release Mechanism used to lock down the PPM at launch \\
$^c$:  PPM 90$^\circ$ measurements made after rotation from 60$^\circ$ to 90$^\circ$ \\
$^d$: PPM 90$^\circ$  measurements made after rotation from 179$^\circ$ to 90$^\circ$ \\
$^e$: Net shift in spin phase measured relative to the position 30$^\circ$ from the Y-axis \\
$^f$:  Net shift measurement from the nominal PPM position 
}
\end{deluxetable*}

\section{Reference Frame Consistency: Ecliptic J2000}
\label{app:frame}

All data products presented in this work utilize a consistent Ecliptic J2000 reference frame for spacecraft attitude, position, and velocity vectors. The Ecliptic J2000 coordinate system is defined with its fundamental plane as the ecliptic plane at the J2000.0 epoch (JD 2451545.0), with the x-axis pointing toward the vernal equinox at that epoch. This frame provides a stable, non-rotating reference that is particularly well-suited for heliospheric science observations, as it naturally aligns with the orbital planes of Solar System bodies and minimizes secular variations in coordinate transformations.

The  data processing pipeline has been checked for rigorous frame consistency throughout the entire analysis chain, from raw star sensor measurements through final science data products. Attitude data, spacecraft ephemeris data, and derived pointing vectors are  consistently expressed in Ecliptic J2000 coordinates. This consistency is critical for accurate determination of the interstellar flow parameters, as even small systematic errors in the reference frame can introduce biases in the derived elongation angles and source directions. We have verified the frame transformations through independent cross-checks and consistency checks. For example, the S/C longitude was compared with the Ecliptic J2000 coordinates of the Lagrangian 1 point. Further, the star sensor stellar positions when compared to known star positions showed consistency in the Ecliptic J2000 frame after proper transformations were applied from the Mean-of-Date reference frame. 

\bibliographystyle{aasjournalv7}
\bibliography{xing_tubes}{}



\end{document}